\documentclass[11pt]{article}
\usepackage[top=23.4mm, bottom=23.4mm, left=23.4mm, right=23.4mm]{geometry}
\usepackage{graphicx,dcolumn,bm}
\usepackage{amssymb,amstext,amsmath}

\usepackage{graphicx}
\usepackage{dcolumn}
\usepackage{bm}
\usepackage{amssymb}
\usepackage{multirow}
\usepackage{bigstrut}
\usepackage{makecell,rotating}
\usepackage{mathrsfs}
\usepackage{booktabs}
\usepackage{threeparttable}
\usepackage{multirow}
\usepackage{subfigure}
\usepackage{colortbl}
\usepackage{epsfig}
\usepackage{threeparttable}
\usepackage{chngpage}
\usepackage{float}
\usepackage{amstext}
\usepackage{amsmath}
\usepackage{amsthm}
\usepackage{pdflscape}
\usepackage{authblk}
\usepackage{hyperref}
\usepackage{rotating}
\usepackage[graphicx]{realboxes}
\usepackage{adjustbox}
\usepackage{setspace}
\usepackage{caption}
\usepackage{xcolor}
\usepackage{xr}
\usepackage{caption}
\usepackage{lineno}
\usepackage{caption}
\usepackage{pdfpages}

\usepackage{enumitem}
\setitemize[1]{itemsep=1pt,partopsep=0pt,parsep=\parskip,topsep=4pt}
\title{Adaptive coordination promotes collective cooperation in repeated social dilemmas}
\author{Feipeng Zhang$^{1\dagger}$, Te Wu$^{1\dagger}$ and Long Wang$^{2*}$}
\date{
    \footnotesize $^1$  Center for Complex Systems, Xidian University, Xi’an, China\\
    $^2$  Center for Systems and Control, College of Engineering, Peking University, Beijing, China\\
    $^{\dagger}$ These authors contributed equally to this work\\
    $^*$ Corresponding author. E-mail: longwang@pku.edu.cn\\[2ex]
}
\begin{document}
\maketitle

\begin{abstract}
    Direct reciprocity based on the repeated prisoner’s dilemma has been intensively studied. Most theoretical investigations have concentrated on memory-$1$ strategies, a class of elementary strategies just reacting to the previous-round outcomes. Though the properties of ``All-or-None" strategies ($AoN_K$) have been discovered, simulations just confirmed the good performance of $AoN_K$ of very short memory lengths. It remains unclear how $AoN_K$ strategies would fare when players have access to longer rounds of history information. We construct a theoretical model to investigate the performance of the class of $AoN_K$ strategies of varying memory length $K$. We rigorously derive the payoffs and show that $AoN_K$ strategies of intermediate memory length $K$ are most prevalent, while strategies of larger memory lengths are less competent. Larger memory lengths make it hard for $AoN_K$ strategies to coordinate, and thus inhibiting their mutual reciprocity. We then propose the adaptive coordination strategy combining tolerance and $AoN_K$' coordination rule. This strategy behaves like $AoN_K$ strategy when coordination is not sufficient, and tolerates opponents’ occasional deviations by still cooperating when coordination is sufficient. We found that the adaptive coordination strategy wins over other classic memory-$1$ strategies in various typical competition environments, and stabilizes the population at high levels of cooperation, suggesting the effectiveness of high level adaptability in resolving social dilemmas. Our work may offer a theoretical framework for exploring complex strategies using history information, which are different from traditional memory-$n$ strategies. 
\end{abstract}

\section*{Introduction}
Group cooperation is one of the cornerstones of human society and is, as such, also intrinsically entwined with the prosperity of our species\cite{Axelrod1981, Axelrod1988, Kerr2004, Rand2013, Kleshnina2023}. However, there is a mismatch between the interest of the individual and of the group. Such social dilemmas\cite{Hauser2019, McAvoy2020} arise at every level of human society and in scenarios ranging from domestic conflicts to global climate change. How has evolution successfully shaped cooperative behaviours in such dilemmas? The cooperative dilemma has been intensively studied for decades and, fortunately, some mechanisms have forestalled us from giving in to the tragedy of non-cooperation\cite{Wang2023, Su2022, Braga2022, Allen2017, Wu2017, Fu2008, PRL2007, Chen2008}. Cooperation may be worthwhile if players realize that they can interact over many rounds. Such a shadow of the future discourages players from cheating on others without hesitation\cite{Nowak1993, Park2022}. Consequently, the pursuit of long-term relationships built on mutual cooperation is more likely to flourish in the population\cite{Segbroeck2012, Hilbenhb2018}. This constitutes the essence of direct reciprocity\cite{Trivers1971}. In this context, win-stay lose-shift ($WSLS$) stands out, which is simply repeating one's own action in the previous round when doing well, and altering otherwise\cite{Matsen2004, Kim2021, Imhof2007, Nowak1993}.

In explaining the enhancing effects of direct reciprocity on cooperation, pairwise interactions are frequently assumed\cite{Li2022, Fischer2013, Press2012, Mathieu2017, Li2014, HAUERT2002}. In stark contrast to the real-life scenarios, human interactions over large groups are ubiquitous\cite{Hauert1997, Hilbe2014, Stewart2016, Wu2018}. Yet, as might be expected, the decision-making becomes significantly more complicated when larger groups of members are involved. Pinheiro et al. found that a simple strategy, All-or-None ($AoN$), performs best in evolutionary settings\cite{Pinheiro2014}. The strategy, $AoN$, cooperates only after one round of coordinated group behaviours. Here the coordinated round means that all group members have adopted the same action in the previous round (all cooperated or all defected). The $AoN$ strategy only takes into consideration of the outcome of the last round, so it belongs to the set of so-called memory-$1$ strategies. Are there more successful strategies if players are allowed to memorize more than one round of outcomes? When players make decisions on the basis of the outcomes of the most recent $n$ rounds, we say they are adopting memory-$n$ strategies. Obviously, the number of memory-$n$ strategies exponentially expands as memory length $n$ increases. For example, there are $65,536$ pure memory-$2$ strategies and approximately $1.94 \cdot 10^{19}$ pure memory-$3$ strategies for a two-person two-action social dilemma. This order of magnitude further increases dramatically when considering social dilemmas involving more than two persons such as in the public goods game. So, it is impossible to explore the entire space of memory-based strategies due to the complexity of the strategies themselves and the prohibitively computational expenses. To advance theoretical analysis, previous studies are mainly concerned with memory-$1$ strategy space. Using an axiomatic approach, Hilbe et al. have intensively explored memory-$n$ strategy space in repeated games. They proposed three key principles that evolutionary stable strategies should bear, upon which they formulated the $AoN_K$ strategies, applicable to multi-person dilemmas as well as players with arbitrary memory length\cite{Hilbe2017}. $AoN_K$ players cooperate only when the actions of all players over the recent $K$ rounds are coordinated. Thus, increasing $K$ means increasing difficulty for $AoN_K$ players to re-coordinate once coordinations are disturbed. In this framework, they found that $WSLS$ is the unique memory-$1$ strategy endowed with the three principles within the context of two-person dilemmas. In fact, both the $WSLS$ strategy and $AoN_K$ strategies exhibit Pavlovian property, which though very simple, plays a very important part in promoting the evolution of cooperation, especially for small and intermediate benefits of cooperation.

However, several essential mathematical features of $AoN_K$ remain largely unaddressed. Though pertinent studies consistently established the evolutionary stability of $WLSL$ strategy in memory-$1$ strategy space, and the good performance of $AoN_2$ strategy as well as strategies equivalent to it, $AoN_K$ strategies have a very obvious weakness. As memory length increases, the efficiency of such strategies would be reduced in noisy environments. This is easily conceived. Imagine a population consisting of all $AoN_K$ strategies. For larger $K$, interactions between such $AoN_K$ strategies would lead to a long period of mutual retaliation once an implementation error occurs, retarding the recovery of mutual cooperation. Thus the population welfare would abate due to the declining cooperation rate. In this paper, to reveal the drawback of $AoN_K$ strategies, we constructed a model to explore the cooperation rate and derive payoffs analytically for $AoN_K$ strategies across arbitrary memory lengths and group sizes. Our results show that the $AoN_K$ strategy achieves the best performance for small-sized groups and intermediate memory length. Moving away from this optimal memory length, the abundance of the corresponding $AoN_K$ strategies declines quickly. So increasing the memory length in $AoN_K$ strategies does not always improve their performance.

A natural question arises as how players make decisions when they have access to a larger amount of history information than just the most recent two-round outcomes? Through which way does direct reciprocity work? Indeed, prior experimental research found that intelligent players as humans use strategies a little more complex than memory-$n$ strategies. Addressing this issue is of critical importance if we are to fully understand the essence of human direct reciprocity, especially in complex scenarios. On the one hand, exploration beyond traditional memory-$n$ strategy space, especially theoretically, is of great necessity to adapt to the rapidly developed artificial intelligence technologies, some of which nowadays are in great lack of explainable theoretical basis. On the other hand, memory-$n$ strategies are insufficient to capture the true intelligence of real intelligent players such as human beings. To better reveal the structure of decision-making of highly intelligent players entails more complex strategy structures. In this work, we propose the adaptive coordination ($ADCO$) strategy which combines tolerance with the coordination rule of $AoN_K$ strategies. It takes advantage of history information to a larger extent. The $ADCO$ player interacting with opponents behaves like a traditional $AoN_K$ strategy, always adopting the action of defection when the coordination level does not reach the threshold, and cooperating when the threshold is reached. Unlike the $AoN_K$ strategy, the $ADCO$ strategy also memorizes how many coordinated rounds have lasted since the first coordination over the last $K$ rounds, called $K$-round coordination. Once the number of $K$-round coordination exceeds a threshold, it tolerates occasional action implementation errors by recounting the number of $K$-round coordination rather than retaliating via defection immediately. The integration of this simple feedback mechanism into the decision-making enhances the $ADCO$'s ability to distinguish opponents. In particular, when opponents have behaved the same way for a long period of rounds, the $ADCO$ player would regard these opponents as co-species and cooperates. At this time, even if one of the opponents occasionally adopts defection, the $ADCO$ player still cooperates but its trust in opponents needs to be rebuilt. Our proposed strategy $ADCO$ overcomes the drawback of the Tit-for-tat ($TFT$) strategy retaliating too quickly, and the drawbacks of $AoN_K$ strategies retaliating too quickly and too many rounds\cite{Nowak2006}. This comes at the price of making use of a larger amount of history information.

Evolutionary game theory provides a powerful mathematical framework to analyze the intricate behaviours of individuals in populations\cite{Hilbe2018, Szolnoki2021, Zhang2023, Wu2023}. Within this framework, we combine theoretical analysis and numerical simulations to evaluate the effectiveness of $ADCO$ strategies in a wide variety of typical population niches. In each of these niches, many classic strategies compete with $ADCO$ strategies to survive in evolutionary races. Results show that the $ADCO$ strategies are more robust than the $AoN_K$ strategies to implementation errors. The enhanced robustness is very crucial in maintaining high levels of cooperation in populations. As is well-established, the ability of a strategy's resistance against the invasion of unconditional defectors is also closely related to its viability. We found that the $ADCO$ strategy has a very good performance in defending defectors' invasion. As the population evolves, defector-defector interactions yield zero payoff. Defectors cannot exploit $ADCO$ strategies, while mutual reciprocity reinforces among the $ADCO$ strategies themselves. As a result, the dominance of $ADCO$ strategies over $ALLD$ becomes more pronounced. The population is eventually pervaded by the $ADCO$ strategies and exhibits very high levels of cooperation. More importantly, results show that the $ADCO$ strategies also stand out in most typical scenarios when one of the classic strategies competes with $ADCO$, and when multiple strategies are simultaneously feasible to players. We hope that our theoretical analysis and the findings could motivate attention focusing on complex strategic structures beyond memory-$K$ strategies and more possible evolving paths towards the emergence and stability of cooperative behaviours.

\section*{Results}
\paragraph{Model description.}
We consider group decisions involving $N$ participants in the framework of repeated social dilemmas. In each round, players independently decide whether to cooperate (C) or defect (D). The prisoner's dilemma is the most popular mathematical metaphor for studying cooperation puzzles when it comes to the two-person dilemma. In this scenario, mutual cooperation (CC) yields a reward $R$, while mutual defection (DD) results in a punishment $P$. If the two players choose different actions (CD or DC), the defector receives the temptation $T$ to defect, while the cooperator receives the sucker's payoff $S$. The payoff parameters satisfy two inequalities $T > R > P > S$ and $T+S<2R$. The constraint $T > R > P > S$  means that defection is the unique Nash equilibrium in the prison's dilemma and defection is also the best reply to the opponent whatever the opponent chooses. The constraint $T+S<2R$ is a widely involved assumption in repeated prisoner's dilemma game. It ensures that mutual cooperation yields a higher payoff than alternating between cooperation and defection in two consecutive rounds. For multi-player dilemmas, the public goods game is the most typical. It is also called multi-player prisoner's dilemma, as this simple model still captures the essence of social dilemma where the strategy maximizing the collective welfare conflicts with the best choice on the individual level\cite{Hilbe2014}. Other classic models include the volunteer dilemma\cite{Hauert2002Volunteer}, the threshold public goods game\cite{Wangjing2009}, and the public goods game with risk.

Repeated games allow players to adapt to the actions of their fellow players over time. In this study, we are mainly concerned with the infinitely repeated games in noisy environments. In such a repeated game, two players play the prisoner's dilemma infinitely many rounds. As the game play unfolds, information on the history of game outcomes accrues. Players therefore can base their actions on past experiences. In this regard, strategies for a repeated game can be extremely complex. The set of history-based strategies expands exponentially. The intriguing complexity and the exponent curse make it almost impossible to analyze the evolutionary dynamics of the competition of all strategies. Indeed, as the capacity of individual memorizing is realistically constrained, it is widely assumed that individuals have finite memory, thus overcoming the potential drawbacks of excessively long recollections. Another realistic factor is also integrated that actions are subject to implementation errors, which is known as the ``trembling hands" effect in economics literature \cite{Nowak2006, Chen2023}. Let us now present the memory strategies (memory-$n$ strategies) of finite-length memory. A player using memory-$n$ strategy makes decisions based on the outcomes of the most recent $n$ rounds. For the repeated prisoner's dilemma, there are $2^{2n}$ possibilities in terms of the game outcomes of the most recent $n$ rounds. A memory-$n$ strategy needs to prescribe the probability of cooperation given each of these $2^{2n}$ outcomes. Therefore, a memory-$n$ strategy $s$ can be written as 
\begin{equation}s=(p_{\substack{CC...CC\\CC...CC}},p_{\substack{CC...CC\\CC...CD}},...,p_{\substack{DD...DD\\DD...DD}})\end{equation}
The vector $s$ has $2^{2n}$ entries with each prescribing the probability of cooperation given the outcome as shown in the subscript. In the subscript, the implemented actions of the focal player are listed in the first row and that of the opponent in the second row. From left to right presents the most past action to the most recent action. We call $s$ a pure strategy if each of these $2^{2n}$ entries takes the value of either $1$ or $0$. Otherwise, we call $s$ a stochastic strategy. To make the strategy complete, one needs to prescribe the actions for the first $n$ rounds where the outcomes of length $n$ have not been produced. Under noisy environments, these first $n$ actions generally have no effects on the long-term payoffs in infinitely repeated games. We therefore refer to a memory-$n$ strategy as described as expression $(1)$. Of memory-$n$ strategies, a class of strategies called $AoN_K$ stands out as such strategies have typical properties such as mutually cooperating, error-correcting, and retaliation for defective opponents, which enable $AoN_K$ to dominate in the evolutionary race. Seemingly, the performance of such $AoN_K$ strategies would be further enhanced as memory length $K$ increases. Under noisy environments, players' actions are subject to implementation errors. That is, when a player intends to cooperate, he would actually implement cooperation with probability $1-\varepsilon$. With probability $\varepsilon$, an error takes place and the actually implemented action is defection. Similarly, players who intend to defect are also subject to implementation errors.

\begin{figure}[!ht]
    \centering
    \includegraphics[width=0.9\textwidth]{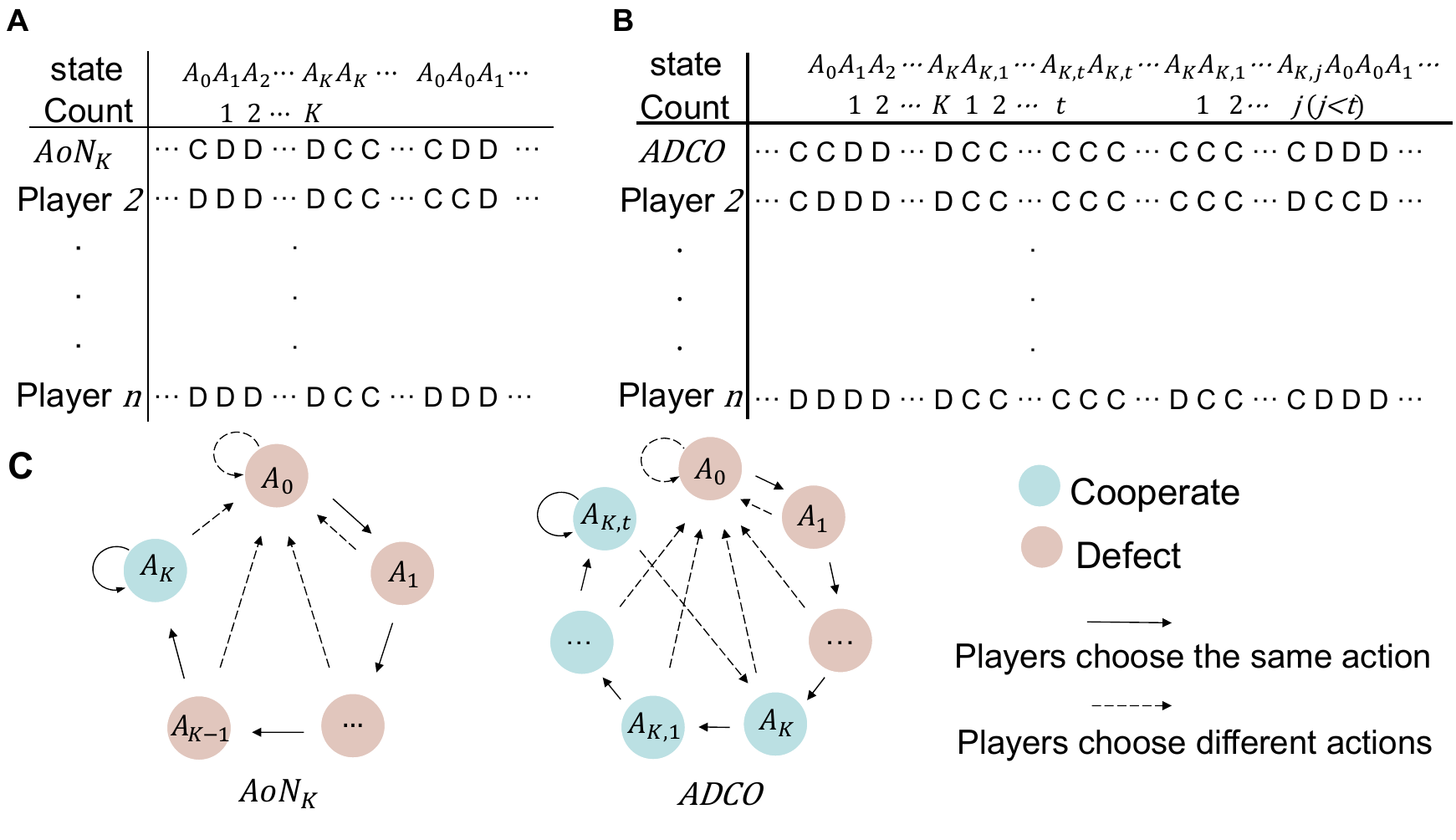}
    \captionsetup{font={small}}
    \caption{\textbf{Illustration of $AoN_K$ strategies and $ADCO$ strategies.} \textbf{(A)} We say the past $I (I=1,2,3,...)$ rounds are coordinated if all players have chosen the same action in each of the previous $I$ consecutive rounds. $AoN_K$ cooperates if the past $K$ rounds are coordinated. Otherwise, it defects. We call $K$  the cooperation threshold. Clearly, $K$ is also the minimum memory length required to implement the $AoN_K$ strategy. Depending on the length of the coordinated rounds so far, the past $K$-round outcomes can be classified as $K+1$ different states $A_i (i = 0, 1, 2,..., K$). $A_i$ represents that the past $i$-rounds are coordinated, and $A_0$ means that the most recent round is uncoordinated. Transitions between these $K+1$ states follow this way. When all players have chosen the same action in the current round, state $A_i$ (when $i < K$) transits to $A_{i+1}$ or stays in this state $A_i$ ($i=K$). When players have adopted different actions in the current round, state $A_i (i=0,1,..., K)$ transits to  $A_0$. Following one round of uncoordinated actions, $AoN_K$ will immediately retaliate for the next $K$ rounds by defecting. \textbf{(B)} For large $K$, an accidental trembling hand can cause $AoN_K$ strategies to fall into a long-term defection, leading to the collapse of cooperation. To explore the nature of direct reciprocity in the context of large memory length, we propose the adaptive coordination ($ADCO$) strategy which incorporates forgiveness into the action-implementation process. An $ADCO$ player with cooperation threshold $K$ and tolerance $t$ behaves the same way as a $AoN_K$ player when less than $K$ uncoordinated rounds emerge. When $K$ coordinated rounds emerge, the $ADCO$ player cooperates and counts how many times state $A_K$ has been maintained continually. When the interactions have stayed in state $A_K$ $t$ times, the $ADCO$ player sets up forgiveness in the sense that it still cooperates when one of the opponents defects but it re-counts. In other states, once the uncoordinated actions occur, the $ADCO$ player defects the next round and the state moves to state $A_0$. \textbf{(C)} State-to-state transitions and decision-making for $AoN_K$ and $ADCO$ strategies. State $A_{k,j}(j\in{1,2,\cdots , t})$ indicates that the state $A_K$ has lasted for $j$ rounds. Following a coordinated round, the state $A_K$ transitions to $A_{K,1}$, states $A_{K,j}$ transition to $A_{K,j+1}$ (where $j < t$), and state $A_{K,t}$ remains in its current state. However, after one round of uncoordinated actions, except for state $A_{K,t}$ transitioning to $A_K$, all other states transition to state $A_0$. Observing additional $t$ rounds, say observation phase, improves $ADCO$'s ability to distinguish opponents, especially between defectors and co-species. For interactions between $ADCO$ players and defectors, the state has little chance to transit to the observation phase. A $ADCO$ player behaves like a $AoN_K$ player and thus effectively eschews the exploitation. For interactions between $ADCO$ players themselves, they are more likely to cooperate many times, triggering their forgiveness. This no doubt enhances the mutual reciprocity between them. The parameter $t$ represents the tolerance level of the $ADCO$ strategy. The smaller $t$ the more forgivable. As $t$ tends to infinity, $ADCO$ with cooperation threshold $K$ is equivalent to $AoN_K$. } \label{fig1}
\end{figure}

\paragraph{$AoN_K$ strategies and $ADCO$ strategies with arbitrary cooperation threshold $K$}
As depicted in Fig. \ref{fig1}, the $AoN_K$ strategies prescribe cooperation only when the outcomes of the preceding $K$ rounds remain coordinated. $K$ is the cooperation threshold for the extent of coordination at which $AoN_K$ players begin to cooperate. The value of $K$ represents the minimum memory length necessary for players to employ the $AoN_K$ strategies. Of course, $AoN_K$ also can be implemented when a player has a larger memory size than $K$. The $AoN_K$ strategy possesses several significant advantages. First of all, it is very effective in maintaining cooperation, as it can consistently achieve full cooperation within a maximum of $K$ rounds, regardless of the current state. Secondly, it is able to correct implementation errors and restore mutual cooperation after at most a maximum of $K$ rounds. In the third place, $AoN_K$ strategies can shun off the exploitation of defectors as such strategies retaliate for at least $K$ rounds if any other member defected in the previous round. Due to this retaliation property, $AoN_K$ strategies can also exploit cooperative strategies such as unconditional cooperation. Except the strategy win-stay-lose-shift, classic memory-$1$ strategies are unable to integrate all these properties due to their very limited responses to just last-round outcomes. For instance, unconditional cooperation can be exploited by defectors. $TFT$ performs quite poorly in handling implementation errors, leading to the alternating between $C$ and $D$ for the interaction between two $TFT$ players. Even very recently discovered $ZD$ strategies are unable to simultaneously achieve both mutual cooperation and resist invasion from defectors\cite{Chen2022, Chennjp2022}.

Previous studies on direct reciprocity have primarily focused on players with cases of small memories, as it simplifies the mathematical tractability. However, such oversimplification may fail to capture the intelligence of real-life agents such as human beings. In this study, we explore the performance of $AoN_K$ strategies with arbitrary cooperation thresholds $K$. Traditionally, the interactions between players using $AoN_K$ are characterized by a Markov chain with finite states. As the memory length increases, the mathematical computations become prohibitively expensive or even impossible. To overcome this limitation, we reconstruct the states and characterize the dynamics of $AoN_K$ strategies' interactions as a finite-dimensional linear system. Of a little surprise, the linear system can be mathematically solved. We are thus able to derive analytical results for numerous significant scenarios (please refer to the Supplementary Information for more details).

\begin{figure}[!ht]
    \centering
    \includegraphics[width=1\textwidth]{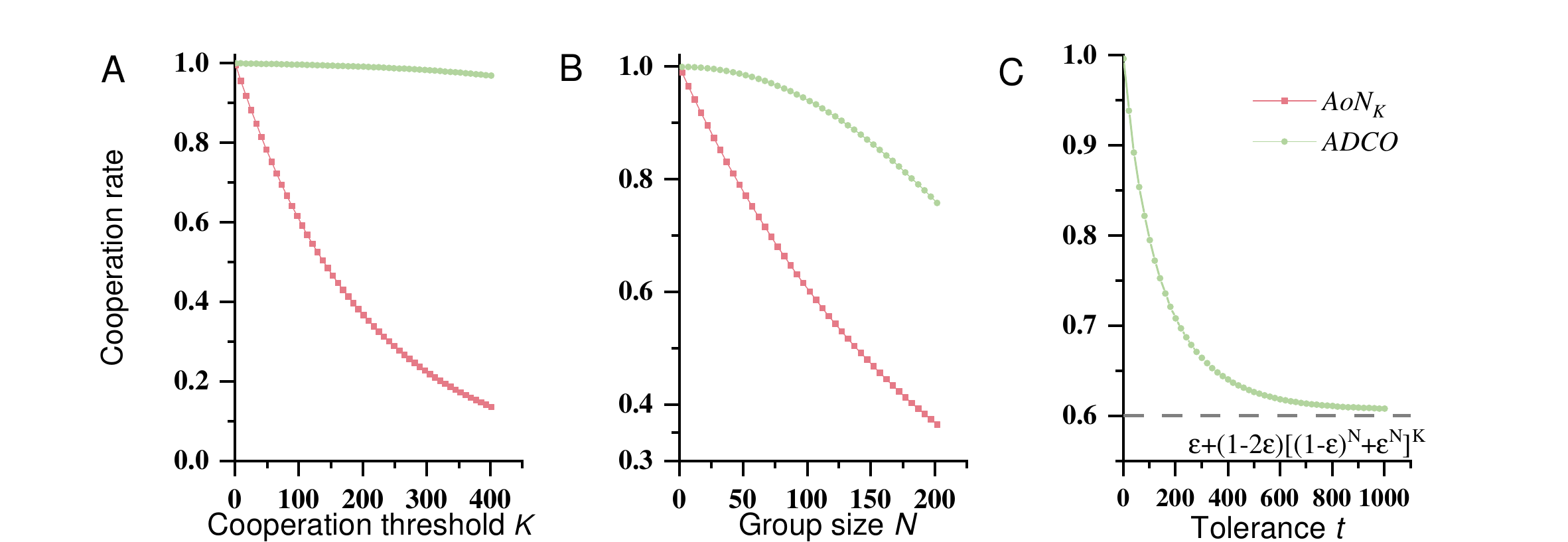}
     \caption{\textbf{Cooperation rate for $AoN_K$ and $ADCO$ strategies for interactions between co-species.} \textbf{(A)} For the same group size, the cooperation rate drastically declines as the cooperation threshold $K$ increases. However, $ADCO$ groups sustain a high level of cooperation for a wide range of $K$.\textbf{ (B) } The cooperation dilemma becomes more challenging as the group size increases. Even so, $ADCO$ players still maintain cooperation level at $>75\%$ for group sizes as large as $200$. However, the cooperation rate dramatically drops as the group size expands. \textbf{(C)} Cooperation rate for different levels of tolerance, where $K=100$. $ADCO$  has limitations in maintaining cooperation when the group size or memory surpasses a certain threshold. A high cooperation rate is difficult to maintain for $ADCO$ of a low level of forgiveness. Parameters in (A) group size $N=5$ and $t=1$; (B) $K=5$ and $t=1$; (C) $K=100$ and $N=5$.} \label{fig2}
\end{figure}

To gain insights into the performance of the $AoN_K$ strategies, we begin by examining games played among $AoN_K$ players. It is evident that the $AoN_K$ strategies can only achieve stable cooperation in self-interaction scenarios. When all players adopt $AoN_K$, the group cooperation rate, denoted as $\rho _{AoN_K}$, can be expressed as follows:
\begin{equation}
    \rho _{AoN_K} =\varepsilon +(1-2\varepsilon)\left [ (1-\varepsilon )^{N} +\varepsilon ^{N}   \right ]^{K},
\end{equation} where $N$ represents the group size. As the frequency of errors $\varepsilon$ decreases, the cooperation rate approaches one. However, as anticipated, the $AoN_K$ strategies fail to sustain cooperation in cases involving large groups and high cooperation thresholds (longer memories). The $AoN_K$ strategies retaliate for at least $K$ rounds when other players choose to defect. While this kind of retaliation can deter selfish players, it also makes the interactions between themselves very sensitive to errors, particularly for $AoN_K$ strategies with longer memories. This no doubt weakens the cooperativeness between $AoN_K$ strategies. Therefore, designing strategies able to shun the error-sensitivity would be of great necessity. Bearing in mind the good performance of high cooperation threshold strategies in resisting invasions by defectors, we design a new kind of strategies that integrate tolerance and the coordination rule of traditional $AoN_K$ strategies. Fig. \ref{fig1} shows the way that the $ADCO$ player plays the repeated games. Unlike traditional memory-$K$ strategies, the $ADCO$ player memorizes how many coordinated rounds consecutively emerged up to now and takes action accordingly. When in state $A_i\left ( i\in\left \{ 0,1,2,\dots, K  \right \}  \right ) $, this player behaves like the $AoN_K$ strategy. Upon the emergence of $K$ coordinated rounds consecutively from state $A_0$, the $ADCO$ player starts to cooperate and records how many coordinated rounds have lasted. For less than $t$ coordinated rounds, any incoordination in any of these rounds would drive the $ADCO$ strategy back to the state $A_0$. Once interactions have yielded $t$ coordinated rounds consecutively, the $ADCO$ player would tolerate a single uncoordinated round but test coplayers again from the state $A_K$. In other words, starting from the state $A_{K,t}$ a single defection of coplayers would drive the $ADCO$ player back to $A_K$ rather than $A_0$. It is obvious that after $t$ rounds of cooperation, the $ADCO$ strategy would tolerate one single defection. The smaller $t$ is, the more tolerant the strategy is. For a group consisting of $N$ $ADCO$ players with cooperation thresholds $K$ and tolerance $t$, the cooperation rate $\rho_{ADCO}$ of the group can be derived as:
\begin{equation}
    \rho_{ADCO} =x_{0} \left [ \frac{1-q^{K}}{1-q}\varepsilon +\frac{q^{K}}{(1-q^t)(1-q)}(1-\varepsilon ) \right ],
\end{equation}
where $q=\varepsilon ^{N} +(1-\varepsilon )^{N}$, and $x_0 = [(1 - q)(1 - q^t)]/[(1 - q^K)(1 - q^t) + q^K]$. Compared with homogeneous groups of $AoN_K$ players, the group cooperation level is remarkably improved for groups of homogeneous $ADCO$ players both for a large range of $K$ and $N$ respectively, verifying the effectiveness of $ADCO$'s tolerance in error-corrections and cooperation coordination. One can also find that high levels of tolerance favor cooperation more (see Fig. \ref{fig2}). When the tolerance level is very low (large $t$), the $ADCO$ strategy behaves in a way very similar to $AoN_K$.

\paragraph{Payoffs against selected strategies.}
Naturally, evaluating a strategy's self-play provides only one perspective on its ability to maintain cooperation. Therefore, we further investigate the pairwise interactions between $AoN_K$ or $ADCO$ strategies and other prominent strategies. Our analysis focuses on the repeated prisoner's dilemma, and the classic game parameters are adopted as in Axelrod's tournaments ($T = 5$, $R = 3$, $P = 1$, and $S = 0$). The magnitude of implementation error is set to be  $\varepsilon =0.1\%$. The payoff of $AoN_K$ against itself, denoted as $\pi (AoN_K,AoN_K)$, can be calculated as follows:
\begin{equation}
   \begin{aligned}
    \pi (AoN_K,AoN_K)=&\left \{\varepsilon ^{2} +(1-2\varepsilon )\left [ \varepsilon ^{2}+ (1-\varepsilon )^{2}\right ]^{K}\right \}R\\
    &+ \left \{   (1-\varepsilon )^{2}-(1-2\varepsilon )\left [ \varepsilon ^{2}+ (1-\varepsilon )^{2}\right ]^{K}\right \}P \\
    &+\varepsilon (1-\varepsilon )(T+S).
    \end{aligned}
\end{equation}

The general procedure used to derive the payoffs for $ADCO$ against itself can be found in the Supplementary Information. Additionally, we analyze the payoffs of the game between different $AoN_K$ strategies and the payoffs of $ADCO$ strategies when playing against any memory-$1$ strategies. We derive the corresponding payoffs by solving a system of linear equations with finite dimensions. Further details can be found in the Supplementary Information.

\begin{figure}[!ht]
    \centering
    \includegraphics[width=0.8\textwidth]{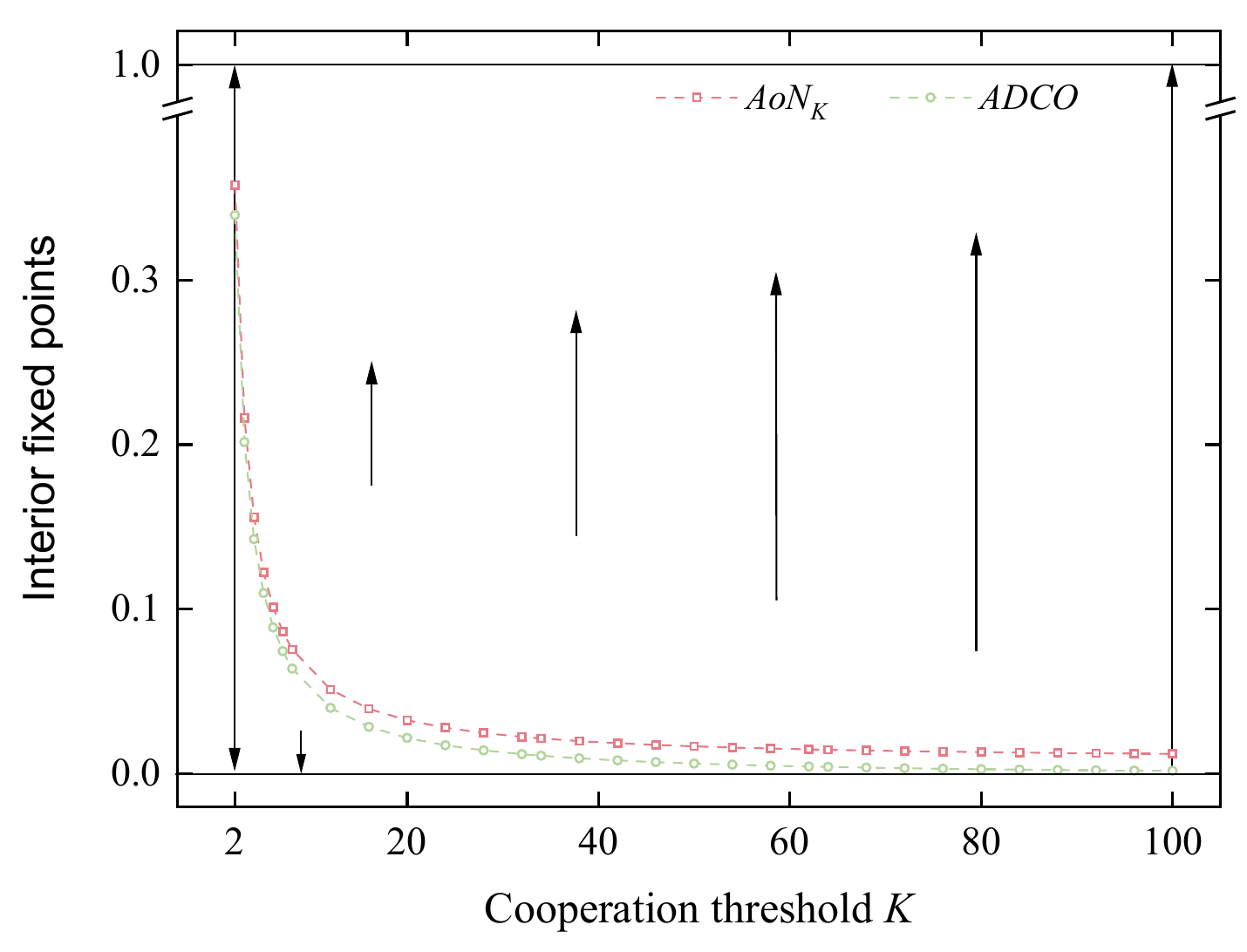}
    \caption{\textbf{Interior fixed points $x_\ast$ of the replicator equation describing the population dynamics of $ADCO$ (or $AoN_K$) competing with $ALLD$.} The population is well-mixed and of infinite size. Denote by $x$ the fraction of $ADCO$ with $t=2$ (or $AoN_K$), and thus $1-x$ represents the fraction of $ALLD$ players. For $K\ge2$, the replicator equation admits three fixed points, two trivial points $x=0$ and $x=1$, and an interior fixed point $x_\ast$. Arrows indicate the direction of selection. The population dynamics exhibit positive frequency dependency. When the initial frequency of $ADCO$ players is above $x_\ast$, the population dynamics would converge to $x=1$. Otherwise, $ALLD$ players would permeate into the whole population. Similar properties are observed when $AoN_K$ competes with $ALLD$ in the population. Increasing the cooperation threshold $K$ improves the ability of $ADCO$ (or $AoN_K$) to resist defectors' invasion as the attraction basin of $ADCO$ (or $AoN_K$) expands, while the improvement in $ADCO$ is more remarkable. } \label{fig3}
\end{figure}

\paragraph{Evolutionary dynamics of two-strategy competition.}
A vital criterion for a cooperative strategy’s success in population is its ability to resist the invasion of unconditional defectors ($ALLD$). So we first consider a well-mixed population of infinite size. The set of strategies available just consists of $ADCO$ and $ALLD$. The strategy evolution in the population is described by the replicator dynamics\cite{Cressman2014}. Denote by $x$ the fraction of $ADCO$ players in the population. The remaining fraction $1-x$ players would be $ALLD$ players. Under replicator dynamics, the change of rate of $x$ in time can be expressed as $\dot{x}=x(f_{ADCO}-\bar{f})$, where $f_{ADCO}$ is the expected payoff of an $ADCO$ player and $\bar{f}$ the average payoff in the population. It can be easily found that the differential equation has two trivial fixed points $x=0$ and $x=1$, and an internal fixed point $x_*$ determined by $f_{ADCO}=f_{ALLD}$. Fig. \ref{fig3} shows the positive frequency dependence of the population dynamics. When the initial frequency of $ADCO$ players is above $x_*$, the population would be stabilized at $x=1$. Otherwise, the population would evolve towards all $ALLD$ players. Similar property is also observed when $AoN_K$ players compete with unconditional defectors in the population. As $K$ increases, the basins of attraction for both strategies significantly expand, suggesting that both $ADCO$ and $AoN_K$ are effective against defectors' invasions. We would like to point out that $ADCO$ is more effective than $AoN_K$ in doing so. On the one hand, the attraction basis for $ADCO$ is slightly larger than that for $AoN_K$. On the other hand, the stable state of the population consisting of $ADCO$ players exhibits a higher level of cooperation than that of $AoN_K$ players. 

\begin{figure}[!ht]
    \centering
    \includegraphics[width=0.9\textwidth,height=1.0\textwidth]{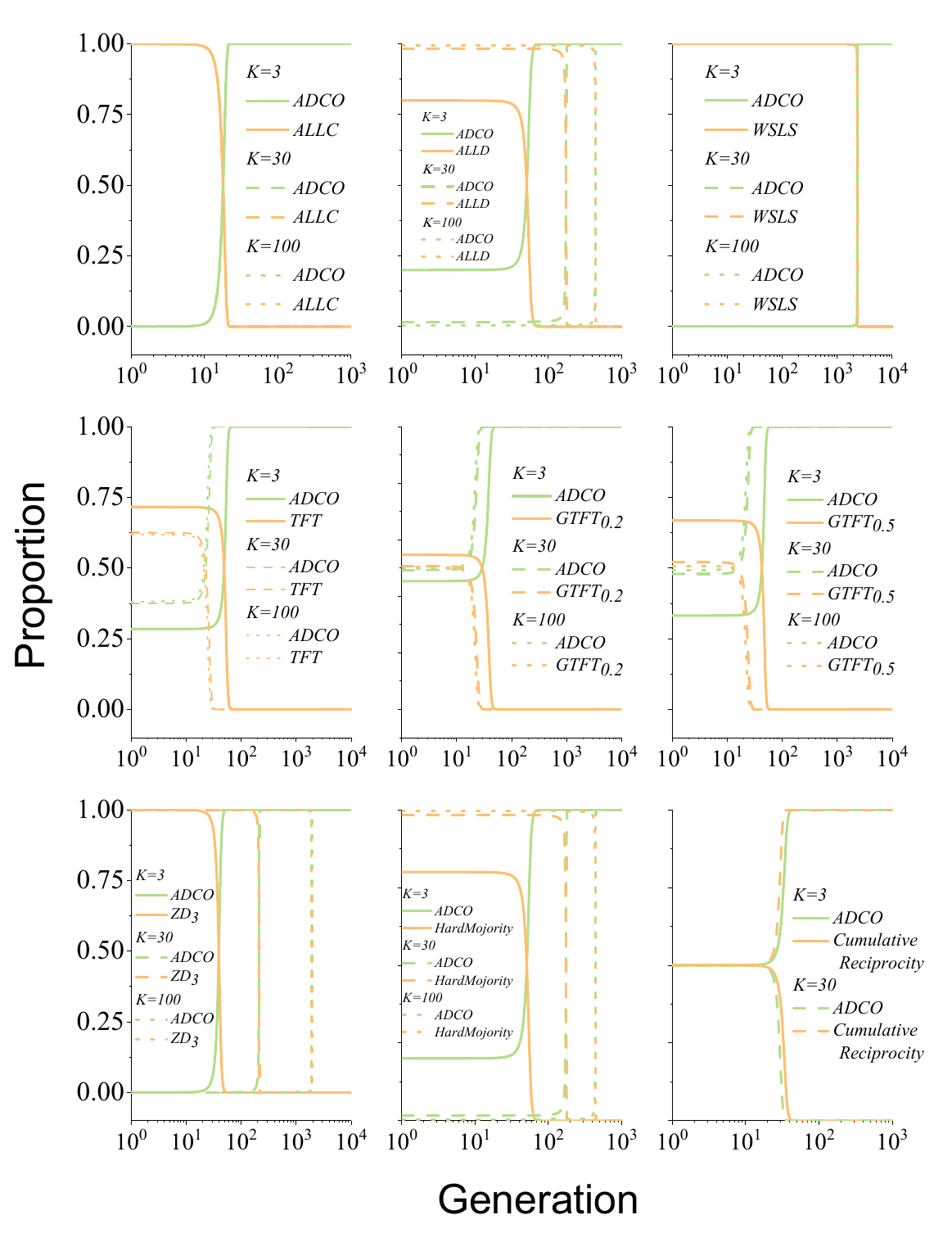}
    \caption{\textbf{Evolutionary dynamics of pairwise competition.}  A large variety of classic strategies are selected to test the evolutionary performances of $ADCO$ $(t=1)$ by vying with each of these classic strategies. Our findings reveal that $ADCO$ exhibits a remarkable ability to invade resident populations all using one type of strategy. It can easily overpower both the cooperative $ALLC$ strategy and the selfish $ALLD$ strategy. Furthermore, $ADCO$ also pervades into the whole population in competitions with tolerant strategies $TFT$, $GTFT_{0.2}$, and $GTFT_{0.5}$, respectively. It is worth noting that the success of $ADCO$ varies over the initial fractions of the resident strategies. For instance, while $ADCO(K=3)$ can invade the resident population of $Cumulative Reciprocity$ ($\Delta=2$) players, $ADCO(K=30)$ cannot. Please refer to the Supplementary Information for details of the strategies involved.} \label{fig4}
\end{figure}

We then examine whether $ADCO$ possesses an evolutionary advantage over other classical strategies in the repeated prisoner's dilemma game. Specifically, we let $ADCO$ compete with each of the widely recognized strategies. To ensure the robustness of our study, we diversify the competition environments by selecting as many classic strategies as possible. The selected strategies include the pure strategies, the stochastic strategies, and complex strategies beyond the description of memory-$K$ strategies. The evolutionary dynamics are described by the replicator dynamics as detailed in the Methods section. We found that the internal equilibrium point does not exist for the differential equations describing the pairwise competitions. As a result, the population would eventually be stabilized in a homogeneous state. Fig. \ref{fig4} demonstrates that $ADCO$ can always pervade into the whole population when in competition with $AllC$, $ALLD$, $GTFT$, $WSLS$, $ZD_3$\cite{Press2012}, $HardMajority$\cite{Mittal2009}, and $Cumulative Reciprocity$ \cite{Li2022}, respectively for most of the parameter sets. Only for $K=30$, $ADCO$ is dominated by the $Cumulative Reciprocity$ strategy. As established, classic memory-$1$ strategies with a certain level of tolerance such as $GTFT$ perform well in many competition settings. However, these strategies are no lack of weaknesses. Let us consider the following simple scenario. The population would neutrally drift when just unconditional cooperators and $GTFT$ compete to survive. Once cooperators dominate, unconditional defectors can expand rapidly. This means that $GTFT$ can be indirectly invaded by defectors. Unlike this indirect evolutionary path, $ADCO$ can not only resist the invasion of defectors but also wins over $GTFT$ directly. These observations illustrate the effectiveness and necessity of combining properties of traditional memory-$K$ strategies in designing strategies in repeated games.

\paragraph{Evolutionary dynamics of multi-strategy competition.}
To further verify the effectiveness of $ADCO$, we consider a larger set of strategies competing in a well-mixed population of finite size $M$. Denote by $S$ the set of strategies available. Following common practice, we use a pairwise comparison process to characterize the population dynamics. The population evolves in discrete time steps. In each evolutionary time step, each player of the population adopts a strategy in $S$. They each interact with all other players and accumulate payoffs. After that, a player (learner) is randomly selected from the population to update the strategy. With probability $\mu$, a mutation happens and the player randomly selects one strategy from the set $S$. With probability $1 - \mu$, learning happens. Then another player is randomly selected from the population as a role model. Denote by $\pi$ and $\tilde{\pi}$ the payoffs of the learner and the model player, respectively. The learner imitates the strategy of the role model with probability $\frac{1}{1+e^{-\beta(\tilde{\pi}-\pi)}}$. The parameter $\beta>0$ represents the intensity of selection, weighing the effects of payoffs on evolutionary dynamics. Its essence lies in that strategies yielding higher payoffs are more likely to spread in the population. For small mutation rates, the population would spend most of the time in homogeneous states where just one strategy takes over the whole population. In this situation, the population dynamics can be approximated by an embedded Markov chain consisting of just these homogeneous states. For larger mutation rates, we use computer simulations to explore the population dynamics.

\begin{figure}[!ht]
    \centering
    \includegraphics[width=\textwidth]{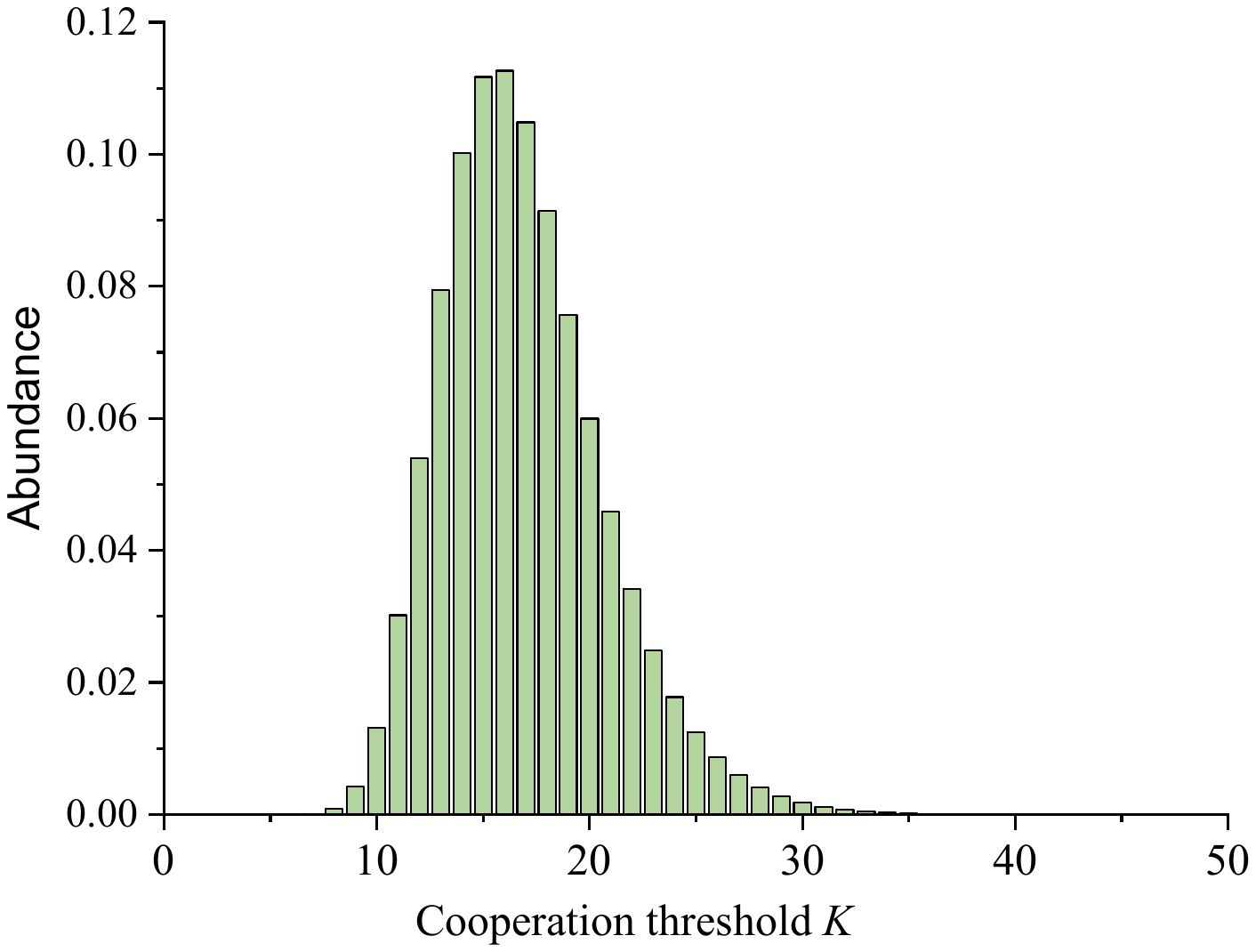}
    \caption{\textbf{Evolutionary dynamics for competitions in the class of $AoN_K$ strategies.} We consider a sufficiently large class of $50$ strategies $(K=1,2,\cdots,50)$. The abundance of strategies is peaked for the intermediate cooperation threshold. Move away from this optimal threshold to either side, the abundance drops. For $AoN_K$ with a low cooperation threshold, they exhibit robust cooperation amidst noise, yet are susceptible to incursion by strategies with high cooperation thresholds. In contrast, $AoN_K$ with a high cooperation threshold can exploit low cooperation thresholds, but interactions among larger $K$ strategies fail to achieve high levels of cooperation, thereby compromising the stability of homogeneous $AoN_K$ populations. Optimal effectiveness for $AoN_K$ strategies is observed at an intermediate cooperation threshold. Specifically, the strategy $AoN_{16}$ is the most abundant in the long run. Parameters: the population size $M=100$, implementation error $\varepsilon=0.01$, and the selection intensity $\beta=1$.} \label{fig5}
\end{figure}

We begin by examining the optimal cooperation thresholds for $AoN_K$ strategies when different $AoN_K$ strategies $(K\in{\left \{1,2,3,...,50\right \}})$ compete in the population. Results show that strategies of varying cooperation thresholds have their respective advantages and weak points. For strategies of low cooperation thresholds, they stabilize the population at very high levels of cooperation in the presence of noise but are vulnerable to invasion of strategies of higher cooperation thresholds. Strategies of high cooperation thresholds can still exploit strategies of shorter memories. However, interactions between the strategies with larger $K$ are unable to achieve high levels of cooperation, weakening the stability of the homogeneous populations of the $AoN_K$ strategy with larger $K$. In between comes a moderate cooperation threshold which proves to be the most effective for $AoN_K$ strategies. This optimal memory length may depend on the specific competition environments including the competing strategies, the noise effect, and the game parameters. As the competition environment is considered now, the strategy $AoN_{16}$ enjoys the largest abundance in the long run (see Fig. \ref{fig5}).

\begin{figure}[!ht]
    \centering
    \includegraphics[width=\textwidth]{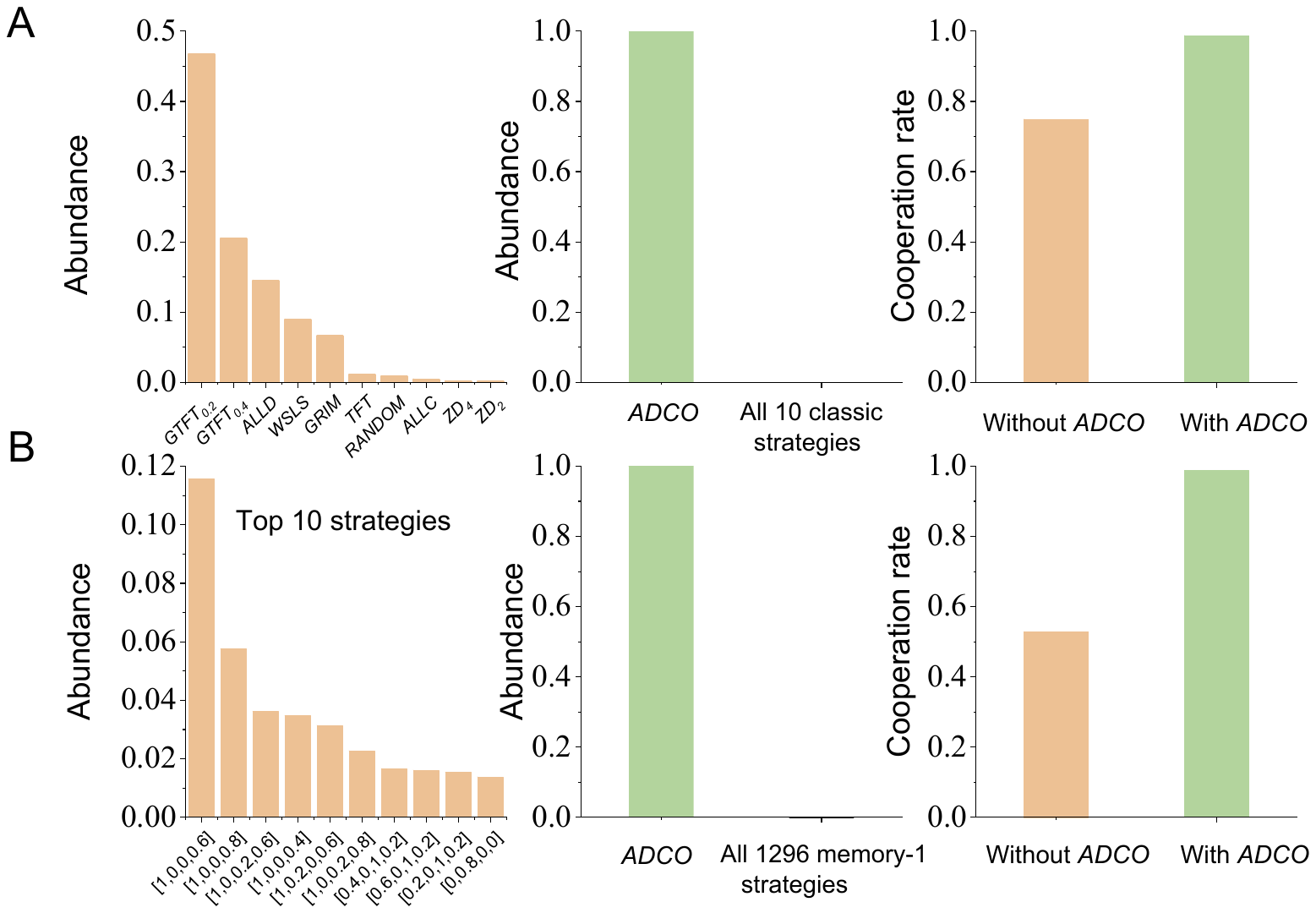}
    \caption{\textbf{$ADCO$'s evolutionary performances in competitions with classic strategies.} \textbf{(A)} We selected $10$ classic strategies (which are described in detail in the main text) to compete with $ADCO$. In the absence of $ADCO$, the generous strategy $GTFT_{0.2}$ enjoys the largest abundance in the long run. The abundances of strategies $GTFT_{0.4}$, $ALLD$, $WSLS$ and $GRIM$ are also non-negligible. When $ADCO$ ($K=3$, $t=2$) takes part in the evolutionary race, it overwhelmingly predominates in the population and thus the abundance of all other strategies combined is vanishingly small. Furthermore, the cooperate rate in the population is also enhanced markedly. \textbf{(B)} In this section, we delved into the evolutionary competition between $ADCO$ and the whole space of  memory-$1$ strategies. Memory-$1$ strategies can be represented as a vector $[P_0, P_{CC}, P_{CD}, P_{DC}, P_{DD}]$ in the 4-dimensional cube $[0, 1]^4$. We select $6\times 6\times 6\times 6$ evenly spaced grid on $[0, 1]^4$. So there are $1296$ memory-$1$ strategies. Similar to the observations in case \textbf{(A)}, in the absence of $ADCO$, a handful of memory-$1$ strategies exhibit quasi-cyclical dominance and their abundances are comparable. Once $ADCO$ enters the competition, the quasi-cyclical dominance is disturbed and $ADCO$ players almost take over the whole population. The cooperation rate is also improved for competition environment with $ADCO$ than without $ADCO$. Parameters: the population size $M=100$, $\varepsilon=0.01$, and the selection intensity $\beta=1$.} \label{fig6}
\end{figure}

To further explore the performance and robustness of $ADCO$, two more complex competition environments are considered (Fig. \ref{fig6}). In the first environment, ten classic strategies $GTFT_{0.4}$, $GTFT_{0.2}$, $WSLS$, $ALLD$, $GRIM$, $ALLC$, $RANDOM$, $TFT$, $ZD_2$, $ZD_4$ are selected to compete with $ADCO$. These strategies are typical as the good performances of some of those strategies have been repeatedly found in respective competition environments\cite{Chen2023}. The selection-mutation dynamics are used to describe the strategy evolution in the population. For small mutation rates, the abundances of strategies correspond to the stable distribution, which can be obtained by solving the eigenvector (associated with the largest eigenvalue) of the embedded Markov chain. In the absence of $ADCO$, $GTFT_{0.2}$ accounts for around $46.8\%$ of the population in the long run. $GTFT_{0.4}$ comes in the second place and its abundance is around $20.5\%$. Though not so abundant, the three strategies $ALLD$, $WSLS$, and $GRIM$ can also take over the population from time to time, and thus their fractions are non-negligible. In addition, extortionate strategies have no chance to evolve, in line with the results reported in \cite{Stewart2013}. The rationale underlying this observation is that the competitions between these strategies exhibit quasi-cyclical dominance. Each strategy can just be effective in competing with some of the ten strategies. In competition with the remaining strategies, this strategy would not be so effective or even be dominated. Using more history information in decision-making can overcome the limitations of memory-$1$ strategies. The strategy $ADCO$ is such a decision-making rule we have proposed. When $ADCO$ enters the competition, the population exhibits qualitatively different dynamics. Fig. 6 shows that in competition with the ten selected strategies, the strategy $ADCO$ enjoys an abundance of nearly $100\%$, indicating that $ADCO$ players can outcompete the selected strategies and take over the population almost all the time. This should be attributed to the integration of advantages in $ADCO$, of previous strategies. Moreover, the cooperation level is around $83\%$ in the absence of the $ADCO$ strategy. With $ADCO$ present, the cooperation level can be improved as high as $100\%$. This means that classic memory-$1$ strategies are unable to invade the population occupied by $ADCO$ players.

In another typical competition environment, we considered a large set of memory-$1$ strategies. Following expression $(1)$, a memory-$1$ strategy can be represented as a vector $[p_0, p_{CC}, p_{CD}, p_{DC}, p_{DD}]$, where $p_0$ represents the probability of cooperating in the first round and $p_{XY}$ represents the player's probability of cooperation when the player has chosen $X\in\{C, D\}$ and the opponent has chosen $Y\in\{C, D\}$ in the previous round. Under a noisy environment, interactions between memory-$1$ strategies would enter stable distributions. Thus $p_0$ has no effect on the long-term payoffs. To simplify our simulations, the values of $p_{XY}$ are constrained to the set \(\{0,0.2,0.4,0.6,0.8,1.0\}\). Consequently, our simulation encompassed a total of $6^4$ memory-$1$ strategies. As a control, we first let these memory-$1$ strategies compete in the evolutionary race. Results are shown in Fig. \ref{fig6}. Although various strategies have a chance to take over the population, the $WSLS$-like strategies are the most prevalent. Due to the high level of diversity of strategies, even the most abundant strategy $[1,0,0,0.6]$ just enjoys a fraction of $11.5\%$ of the time, further corroborating the ebbing effectiveness of each memory-$1$ strategy in interacting with diverse memory-$1$ strategies. We then incorporate the $ADCO$ strategy into the competition environment.  As mentioned above, the $ADCO$ player is tolerant but to a limited degree. This tolerance has a remarkable effect in improving payoffs when interactions happen between $ADCO$ players themselves. Fig. \ref{fig6} shows that the $ADCO$ strategy does have a good performance as it can wipe out all other memory-$1$ strategies and take over the population almost all the time. The abundance of all memory-$1$ strategies combined is vanishingly small. The abundance of $ADCO$ is above $99\%$. In the absence of $ADCO$, the cooperation level in the population is further reduced as the competing strategies diversify. This is understandable. Some strategies can achieve high cooperation levels, while others cannot. The population exhibits quasi-cyclical dominance. On average, the cooperation level in the population is discounted. Once $ADCO$ is introduced to the population, it can wipe out all memory-$1$ strategies. The quasi-cyclical dominance disappears. The population exhibits a very high level of cooperation, as interactions between $ADCO$ players can achieve mutual reciprocity effectively due to the inherent tolerance of the $ADCO$ strategy. 

Using computer simulations, we have also explored the population dynamics for larger mutation rates. For the competition between the family of $AoN_K$ strategies of varying $K$, the $AoN_K$ strategies of intermediate memory length prevail (see Fig. S1). In competing with classical memory-$1$ strategies or a larger set of memory-$1$ strategies, $ADCO$ strategy still dominates in the population for both competition environments (see Fig. S2 and Fig. S3). These findings are qualitatively similar to the ones found under weak mutation, corroborating the robustness of $ADCO$'s good performance in a large range of mutation rates. 

We would like to emphasize that the tolerance in $ADCO$ strategies entails a certain level of risk of being exploited. For instance, players can exploit $ADCO$ players by defecting during a stable cooperative state while cooperating in a non-stable cooperative state. To implement such exploitation, players require an equal or greater amount of history information than $ADCO$ do for decision-making in playing the repeated prisoner's dilemma game. Nevertheless, even if such an exploitation is successful, the increment of cooperation thresholds of the $ADCO$ strategy reduces the possibility of being exploited and thus enhances the evolutionary competence of the $ADCO$ strategy. In fact, in these competition scenarios as considered now, no strategies can observe $ADCO$'s tolerance and exploit it. As a result, the population can always be stabilized at high levels of cooperation when $ADCO$ players take part in evolutionary races. 

\section*{Discussion}
Repeated social dilemmas and conditional cooperation serve as crucial factors for understanding human cooperative behaviours\cite{Imhof2010, Boerlijst1997}. Searching for effective strategies in repeated prisoner's dilemma is still a big challenge in the field of evolutionary game theory. As the game is repeatedly played between the same players, they are able to collect history information and thus plan their actions based on such history information. Due to realistic computational considerations, previous studies along this line have commonly assumed that individuals just base their actions on the outcomes of the most recent $n$ rounds hitherto. The action implementations following such principle are called memory-$n$ strategies. Memory-$1$ strategies are intensively studied primarily due to computational simplicity and easy interpretations\cite{Yi2017, Glynatsi2020, Veelen2016, Baek2016, Matthijs2018, Schmid2022}. Microscopic properties of these strategies have been revealed such as the weakness of $TFT$'s error correction and $ZD$'s enforcement of payoff relationships. Using simulations on larger sets of memory-$2$ (or $3$) strategy space, Hilbe et al. \cite{Hilbe2017}) found that successful strategies have three key properties of $AoN_K$ strategies. These properties include self-cooperation, error correction, and robust retaliation against selfish defectors. However, the performance of $AoN_K$ strategies of larger memory lengths remains unknown. In this work, we revealed an obvious weakness of $AoN_K$ strategies with larger $K$. We then combine the desirable properties of $AoN_K$ with tolerance to generate powerful and cooperative strategies, $ADCO$, that enhance competitiveness and reduce the risk of being invaded.

First of all, we rigorously derived the payoff expressions for a group of players using $AoN_K$ strategies, which allows us to explore the evolutionary dynamics among the competitions of $AoN_K$ strategies of varying cooperation thresholds. In such a competition scenario, the $AoN_K$ strategy with intermediate cooperation thresholds is most abundant. Moving away from the optimal cooperation thresholds on either side, the abundance of the corresponding strategy declines monotonically. Underlying rationales differ. For $AoN_K$ strategies with very small $K$s, they can achieve mutual reciprocity through quick synchronization. For this reason, these strategies are also easily exploited. They have little chance of prevailing in the population. When $AoN_K$ strategies tend to synchronize through too many rounds of defection, they can effectively eschew the exploitation of opponents. The increased defensibility against exploitation also weakens the mutual reciprocity among the players all using such $AoN_K$ strategies. Therefore, they can occasionally prevail and even take over the population. But they cannot stabilize the population for long periods of time in the evolutionary race. As a trade-off, the $AoN_K$ strategies with intermediate $K$s can overcome, to some degree, the obvious weaknesses of strategies with extreme $K$s.

This coordination rule, ``cooperate when synchronized, otherwise defect", enables rapid establishment of stable cooperation when players all follow this rule to make decisions. The well-known memory-$1$ strategy, win-stay, lose-shift ($WSLS$) is such a typical strategy in the repeated prisoner's dilemma. Previous studies have repeatedly demonstrated the good performance of $WSLS$ in the memory-$1$ strategy space. However, due to the restrictions on memory length, the power of this coordination rule has not been fully revealed. Using more information to achieve synchronization proves more effective in identifying interacting co-players, as our findings show. This identification mechanism enables $AoN_K$ players to ward off the exploitation of opponents not using $AoN_K$, to reinforce mutual reciprocity among $AoN_K$ players, two crucial properties of $AoN_K$ strategies in constructing a cooperative society. In this regard, our work expands the scope of study on memory-based strategies in repeated games.

Previous studies have established that using a larger amount of history information for decision-making favors the evolution of cooperation. However, our results reveal the existence of the optimal cooperation thresholds when $AoN_K$ strategies of varying $K$ compete to survive in an evolutionary race. This stark contrast means that the ways of using history information matter. This motivates us to propose a new kind of strategy, $ADCO$, which plans actions based on history information in a way different from the traditional $AoN_K$ strategy. It first observes whether it has coordinated with interacting co-players a sufficient number of rounds. If not, it behaves like the traditional $AoN_K$ strategy and adopts the action of defection. Once they have coordinated enough rounds as expected, the $ADCO$ strategy begins to cooperate. Meanwhile, it also counts how many rounds they have coordinated consecutively from then on. When the coordinated state has been maintained more than a threshold, the $ADCO$ strategy exhibits forgiveness. That is, when some co-player defects, the $ADCO$ strategy still cooperates while returning to the state where they first coordinated $K$ rounds. The $ADCO$ strategy does not belong to the traditional memory-based strategy space, as its memory depth is indefinite. Though this seemingly complexity of the $ADCO$ strategy, we are still able to analytically derive the payoffs when it interacts with classic memory-$1$ strategies. This is convenient for us to probe the performance of $ADCO$ strategies in a large variety of competition scenarios.

When interactions just take place among co-species, the population full of $ADCO$ players sees a significantly higher cooperation rate than the population full of $AoN_K$ players. This enhances the evolutionary stability of the $ADCO$ strategy. When competing with $ALLD$ players, the $ADCO$ players also perform better than $AoN_K$ players, as the attraction basin of the $ADCO$ players is generally larger. Using ecological dynamics, we have corroborated that $ADCO$ players can win over classic strategies including $ALLC$, $ALLD$, $WSLS$, $TFT$, $GTFT$, $ZD$, and $HardMajority$ in pairwise competitions. In addition, $ADCO$ players can also beat the $Cumulative Reciprocity$ strategy, a well-performed strategy most recently discovered in repeated prisoner's dilemma. Furthermore, when a large set of classic memory-$1$ strategies compete in an evolutionary race, the $[1,0,0,0.6]$ strategy does dominate in the population. However, the abundance of some other strategies is non-negligible and sometimes even comparable. When $ADCO$ strategy enters the competition, it overwhelmingly beats over all memory-$1$ strategies and takes over the population $99\%$ of the time or more, and accordingly the cooperation rate is also improved. Therefore, using a larger amount of information properly can improve the efficiency of decision-making in repeated games. History information storage and memorizing is, at least cognitively, costly. How to integrate the cost into the game payoff awaits future experimental study.

In conclusion, both $AoN_K$ and $ADCO$ emerge as straightforward yet effective cooperation-promoting strategies. While the mathematical analysis of longer-memory strategies may entail some complexity, the implementation of $AoN_K$ and $ADCO$ strategies themselves is relatively straightforward. Hilbe et al. confirmed that strategies akin to $AoN_2$ are the most abundant of memory-$2$ strategies. Our findings show that strategies of intermediate cooperation thresholds are preferred over other strategies with too low or too high thresholds. This means that increasing cooperation thresholds does not always favor the evolution of corresponding $AoN_K$ strategies. In fact, as an $AoN_K$ strategy characterizes an opponent’s similarity by the memorized outcomes of the most recent $K$ rounds, it is greatly vulnerable to implementation errors. When an error has happened, it is carried forward through memory and impedes players' coordination synchronously. This obvious shortcoming weakens the mutual reciprocity between $AoN_K$ players. We proposed the strategy $ADCO$, which combines the typical properties of $AoN_K$ strategy and tolerance. The $ADCO$ player defects when not coordinating with opponents. When it coordinates with opponents sufficiently, it forgives an opponent's occasional defection by still cooperating. Thus the $ADCO$ strategy has a greater ability to distinguish among various types of opponents, especially between highly defective opponents and highly cooperative opponents, and hence may respond differently to identical actions implemented by different opponents. The upgraded adaptive response enables $ADCO$ players to ward off the exploitation of defectors and achieve high levels of mutual reciprocity between themselves. As a result, $ADCO$ strategies perform best in various typical competition scenarios. We hope our work could shed light on better understanding and exploring more possible evolutionary paths toward the emergence and maintenance of cooperative behaviours.

\section*{Methods}
In this section, we provide a comprehensive overview of the setup of evolutionary processes. For pairwise competitions, the population dynamics are characterized by the replicator dynamics, which can clearly reveal the relative dominance of one strategy over another. For multi-strategy competitions, the population dynamics are characterized by the standard imitation dynamics. The imitation process is also subject to mutation. These selection-mutation dynamics can better reveal the evolutionary performances of strategies in confrontation with complex competition environments. 

\paragraph{Payoff calculation.} First of all, we present the procedures for calculating the payoffs for interactions between  $ADCO$ players and between $AoN_K$ players, respectively. For the class of $ADCO$ strategies and the class of $AoN_K$ strategies, we can rigorously derive the payoffs by solving a system of finite linear equations. Furthermore, similar approaches can be used to obtain the payoffs for interactions of $ADCO$ against memory-$1$ strategies and of $AoN_K$ strategies against memory-$1$ strategies. For these cases, the number of linear equations is increased but is still finite.  To determine the payoffs for interactions between memory-$1$ strategies, we utilize the analysis method outlined in ref.\cite{Braga2022}. For the remaining scenarios, we use computer simulations to estimate payoffs.

\paragraph{Replicator dynamics for pairwise competitions.} We use the replicator dynamics to characterize the change of the proportions of strategies in the evolutionary process. Consider a well-mixed population of infinite size. Each individual adopts one of two competing strategies $s_i$ and $s_j$. Every individual has the same chance to interact with every other individual in the population. In the $t$th generation, denote by $x_i(t)$ the proportion of strategy $s_i$ and thus $1-x_i(t)$ the proportion of strategy $s_j$ in the population. The payoff of an $s_i$ individual is expected to be $f(s_i)=x_i(t)\pi (s_i, s_i)+(1-x_i(t))\pi (s_i,s_j)$, where $\pi (s_i,s_j)$ represents the payoff of strategy $s_i$ against strategy $s_j$. The average payoff of the population is $\bar{f} = x_i(t) f(s_i) + (1 - x_i(t)) f(s_j)$. At the end of each generation, the accumulated payoffs of each strategy are transformed into a new proportion of individuals that adhere to the same strategy during the next generation. Therefore, we can write down $x_i(t+1)=\frac{f(s_i)}{\bar{f}}x_i(t)$. This process simulates selection pressures in the sense that successful individuals would have a larger proportion as evolution advances. This fundamental evolutionary process is repeated until the population is equilibrated. In the equilibrium states, either one strategy takes over the whole population or the proportions of the strategies do not change over time anymore.  

\paragraph{Full population dynamics for competitions between $ADCO$ and memory-$1$ strategies.} To further test the performances of $ADCO$, we consider two typical competition environments admitting $ADCO$ and memory-$1$ strategies. In the first environment, we let $ADCO$ compete with multiple classic memory-$1$ strategies. In another environment, we further diversify the types of competing strategies by adding $6^4$ memory-$1$ stochastic strategies into the environment. These stochastic strategies are sampled unbiasedly from the memory-$1$ strategy space. Following common practice, we use a standard imitation process to describe how strategies evolve over time in the population. Consider a well-mixed population of finite size $(=M)$. At each evolutionary time step, every individual interacts with all other individuals and accrues a payoff. Then, two individuals, a learner and a role model, are randomly selected from the population. With probability $\mu$, mutation takes place. The learner explores the set of feasible strategies by randomly picking one from it. With the supplementary probability $1-\mu$, learning happens. In this case, the learner adopts the strategy of the role model with probability ${[1+e^{-\beta(\tilde{\pi}-\pi)}]}^{-1}$, where $\pi (\tilde{\pi})$ is the payoff of the learner (the role model), and the selection intensity $\beta$ measures the impact of game payoffs on the viability of strategies. For $\beta=0$, payoffs have no effect on evolutionary dynamics and the population drifts neutrally. For $\beta>0$, strategies yielding higher payoffs are more likely to spread in the population.

Generally speaking, it is difficult to rigorously analyze the population dynamics. The agent-based computer simulations are the main techniques to explore the performances of strategies in the evolutionary processes. However, in the limit of rare mutations, the population most of the time admits at most two competing strategies simultaneously in the evolutionary process. As the mutation events are so rare when a mutant occurs, either it invades the resident population and fixates, or it is eliminated before the next mutant appears. Therefore, for small $\mu$, the selection-mutation dynamics can be approximated by an embedded Markov chain consisting of finite states, each corresponding to one homogeneous population occupied by one strategy. The transition rate $T_{ij}$ from state $i$ to state $j$ is the fixation probability that $j$ as mutant invades successfully the resident population $i$ and takes over the whole population. Thus, $T_{ij}$ can be written as 
\begin{equation*}
    T_{i,j} =\left[1+\sum_{k=1}^{N-1} \prod_{l=1}^{k} e^{-\beta\left[\pi_{M}(l)-\pi_{R}(l)\right]}\right]^{-1},
\end{equation*} 
where $\pi_{M}(l)$ and $\pi_{R}(l)$ are the payoffs of the mutation strategy and the resident strategy respectively when there are $l$ mutants in the population. The parameter $\beta$ is the selection intensity. In this way, we get the non-diagonal entries of the transition matrix $T$. The diagonal entry in each row can be obtained as one minus all other entries in this row. We can then derive the left eigenvector of the transition matrix $T$ corresponding to its largest eigenvalue of one. This left eigenvector, after normalization as needed, represents the stationary distribution of these finite states. Using this stationary distribution as strategy abundance, the cooperation level in the population can also be estimated by letting each of these strategies interact with every other strategy and itself.

\bibliography{references}

\clearpage
\newpage



\section*{Supplementary material}

# Adaptive coordination promotes collective cooperation in repeated social dilemmas

Feipeng Zhang, Te Wu, Long Wang

In the following, we present a complete description of the mathematical and numerical methods employed to obtain the outcomes presented in the main body of the text. Firstly, we present a mathematical analysis of the cooperation rate and payoff for the $AoN_K$ strategy and $ADCO$ strategy in the context of self-play, for arbitrary cooperation thresholds $K$. Assessing a strategy's performance through self-cooperative maintenance is crucial, particularly in the presence of error perturbations, as it signifies the ability of a cooperative strategy to effectively sustain cooperation. Furthermore, it is noteworthy to examine how this strategy performs against a variety of other strategies. On the one hand, we derive the expression for the payoff in the game between any two distinct $AoN_K$ strategies. on the other hand, We obtain the payoff of the $AoN_K$ (or $ADCO$ with cooperation threshold $K$ and tolerance $t$) strategy versus any of the memory-1 strategies by solving a $2K + 2$ (or $2(K + t) + 4$)-dimensional system of linear equations. Additionally, numerical simulations are utilized to determine the game payoffs for other strategies mentioned in the paper. Finally, we provide a detailed description of the evolutionary dynamics within the populations described in the main body of the paper. This description elucidates how the strategies evolve and interact with each other in the population over time, shedding light on the dynamics and outcomes observed in the study.

## 1 Mathematical Analysis of $AoN_K$ and $ADCO$ Strategies for Arbitrary Memory.

Consider an infinitely repeated social interaction in a group of $N$ players. Players decide simultaneously each round to cooperate (C) or defect (D). $AoN_K$ players will cooperate if all members behaved the same in the past $K$ rounds, otherwise, they will defect. So, $K$ is the minimum memory length required by the strategy. $AoN_K$ is very sensitive to error when $K$ is large. To improve the cooperative robustness of $AON_K$ under long memory, we propose the $ADCO$ strategy. It is reasonable to be tolerant of or ignore occasional errors when players have cooperated for a certain number of consecutive rounds. If player behavior is uncoordinated in the current round, instead of immediately retaliating for round $K$ as $AoN_K$ does, $ADCO$ will tolerate once (continue to cooperate) if all players have cooperated in the past $t$ rounds. $t$ is the tolerance of the strategy, and the smaller it is, the more tolerant it is.

**Proposition 1.** *[AoN$_K$ vs. AoN$_K$] Consider an infinitely repeated social dilemma in a group of $N$ players. Let $\varepsilon > 0$ represent the probability of a player making an error in any given round. Suppose all $N$ players use the AoN$_K$ strategy. Then, the group's average cooperation rate is*

$$\rho_{AoN_K} = \varepsilon + (1 - 2\varepsilon)\left[(1-\varepsilon)^N + \varepsilon^N\right]^K. \tag{S1}$$

*In the case of the prisoner's dilemma with possible payoffs $R$, $S$, $T$, and $P$, the payoff for each player can be expressed as:*

$$\begin{aligned}\pi(AoN_K, AoN_K) &= \left\{\varepsilon^2 + (1-2\varepsilon)\left[\varepsilon^2 + (1-\varepsilon)^2\right]^K\right\} R \\ &+ \left\{(1-\varepsilon)^2 - (1-2\varepsilon)\left[\varepsilon^2 + (1-\varepsilon)^2\right]^K\right\} P \\ &+ \varepsilon(1-\varepsilon)(T+S).\end{aligned} \tag{S2}$$

*Proof.* We categorize the outcomes of the past $K$ steps into $K+1$ states based on the number of latest consistent rounds, denoted as $A_0, A_1, A_2, \cdots, A_K$. $A_i$ means that all players chose the same action in the most recent $i$ rounds. The state $A_0$ means that the most recent round of players' actions was inconsistent. We only focus on the last $K$ steps so the maximum value of $i$ is $K$. $x_i$ denotes the probability that the player is in state $A_i$ in the repeated game. Let $q = \varepsilon^N + (1-\varepsilon)^N$, then we get

$$x_i = qx_{i-1}, i = 1, 2, \cdots, K-1,$$
$$x_K = qx_{K-1} + qx_K.$$

Since all probabilities must end up being one, so

$$\sum_{i=0}^{K} x_i = 1.$$

We get

$$x_i = (1-q)q^{i-1}, i = 1, 2, \cdots, K-1,$$
$$x_K = q^K.$$

Now, let's calculate the average cooperation rate of the players. By definition, in states 0 to $K-1$, the player's cooperation rate is $C_N^N \varepsilon^N + \frac{N+1}{N} C_N^{N-1} \varepsilon^{N-1}(1-\varepsilon) + \cdots + 0 \cdot C_N^0 (1-\varepsilon)^N$ and in state $K$

2

to infinity, the player's cooperation rate is $C_N^N(1-\varepsilon)^N + \frac{N-1}{N}C_N^{N-1}(1-\varepsilon)^{N-1}\varepsilon + \cdots + 0 \cdot C_N^0 \varepsilon^N$. So

$$\begin{aligned}\rho_{AoN_K} &= \sum_{i=0}^{K-1} x_i \left[ C_N^N \varepsilon^N + \frac{N-1}{N} C_N^{N-1} \varepsilon^{N-1}(1-\varepsilon) + \cdots + 0 \cdot C_N^0 (1-\varepsilon)^N \right] \\ &+ x_K \left[ C_N^N (1-\varepsilon)^N + \frac{N-1}{N} C_N^{N-1}(1-\varepsilon)^{N-1}\varepsilon + \cdots + 0 \cdot C_N^0 \varepsilon^N \right] \\ &= x_0 \frac{1-q^K}{1-q}\varepsilon + x_K(1-\varepsilon) \\ &= \varepsilon + (1-2\varepsilon)\left[(1-\varepsilon)^N + \varepsilon^N\right]^K \end{aligned}$$

We use an identical approach to determine player payoffs in the prisoner's dilemma, resulting in the equation

$$\begin{aligned}\pi(AoN_K, AoN_K) &= \left\{ \varepsilon^2 + (1-2\varepsilon)\left[\varepsilon^2 + (1-\varepsilon)^2\right]^K \right\} R \\ &+ \left\{(1-\varepsilon)^2 - (1-2\varepsilon)\left[\varepsilon^2 + (1-\varepsilon)^2\right]^K \right\} P \\ &+ \varepsilon(1-\varepsilon)(T+S).\end{aligned}$$

□

**Proposition 2.** *[ADCOvs.ADCO] Follow the assumption of Proposition 1. Suppose all N players use the ADCO strategy with cooperation threshold K and tolerance t. The average cooperation rate of the group is*

$$\rho_{ADCO} = x_0 \left[ \frac{1-q^K}{1-q}\varepsilon + \frac{q^K}{(1-q^t)(1-q)}(1-\varepsilon) \right], \tag{S3}$$

*where $q = \varepsilon^N + (1-\varepsilon)^N$ and $x_0 = \frac{(1-q)(1-q^t)}{(1-q^K)(1-q^K)+q^K}$. In the two-person prisoner's dilemma, each player's payoff is*

$$\begin{aligned}\pi(ADCO, ADCO) = x_0 &\left\{ \left[\frac{1-q^K}{1-q}\varepsilon^2 + \frac{q^K}{(1-q^t)(1-q)}(1-\varepsilon)^2\right] R \right. \\ &+ \left[\frac{1-q^K}{1-q}(1-\varepsilon)^2 + \frac{q^K}{(1-q^t)(1-q)}\varepsilon^2\right] P \\ &\left. + \varepsilon(1-\varepsilon)(T+S)\right\}.\end{aligned} \tag{S4}$$

*Proof.* On the basis of Assumption 1, we add $t$ states $(A_{K,1}, A_{K,2}, \cdots, A_{K,t})$. We use $x_{K+i}$ ($i = 1, 2, \cdots, t$) to denote the probability that a player is in state $A_{K,i}$. According to the *ADCO* rules, the probability $x_i$ satisfies the following equation:

$$\begin{aligned}x_i &= q \cdot x_{i-1}, i = 0, 1, \cdots, K \\ x_i &= q \cdot x_{i-1}, i = K+1, , K+2, \cdots, K+t-1, \\ x_K &= q \cdot x_{K-1} + (1-q)x_{K+t}, \\ x_{K+t} &= q x_{K+t-1} + q x_{K+t}.\end{aligned}$$

3

Because the probabilities sum to one,

$$1 = \sum_{i=0}^{K+t} x_i = x_0 \frac{1-q^K}{1-q} + \frac{x_K}{1-q}.$$

By solving the above equations, we get

$$x_i = \begin{cases} \frac{q^i(1-q)(1-q^t)}{(1-q^K)(1+q^t)+q^K} & i = 0, 1, \cdots, K-1 \\ \frac{q^i(1-q)(1-q^t)}{(1-q^K)(1-q^t)^2+(1-q^t)q^K} & i = K, K+1, \cdots, K+t-1 \\ \frac{q^{K+t}(1-q^t)}{(1-q^K)(1-q^t)^2+(1-q^t)q^K} & i = K+t. \end{cases}$$

Using these probabilities, we can calculate the player's average cooperation rate as:

$$\rho_{ADCO} = \sum_{i=0}^{K-1} x_i \left[ C_N^N \varepsilon^N + \frac{N-1}{N} C_N^{N-1} \varepsilon^{N-1}(1-\varepsilon) + \cdots + 0 \cdot C_N^0 (1-\varepsilon)^N \right]$$
$$+ \sum_{i=K}^{K+t} x_i \left[ C_N^N (1-\varepsilon)^N + \frac{N-1}{N} C_N^{N-1}(1-\varepsilon)^{N-1}\varepsilon + \cdots + 0 \cdot C_N^0 \varepsilon^N \right]$$
$$= x_0 \frac{1-q^K}{1-q}\varepsilon + x_K \frac{1-\varepsilon}{1-q}$$
$$= x_0 \left[ \frac{1-q^K}{1-q}\varepsilon + \frac{q^K}{(1-q^t)(1-q)}(1-\varepsilon) \right].$$

When it comes to the two-person prisoner's dilemma, the player's payoff is

$$\pi(ADCO, ADCO) = \sum_{i=0}^{K-1} x_i \left[ (1-\varepsilon)^2 P + \varepsilon^2 R + \varepsilon(1-\varepsilon)(T+S) \right]$$
$$+ \sum_{i=K}^{K+t} x_i \left[ (1-\varepsilon)^2 R + \varepsilon^2 P + \varepsilon(1-\varepsilon)(T+S) \right]$$
$$= x_0 \left\{ \left[ \frac{1-q^K}{1-q}\varepsilon^2 + \frac{q^K}{(1-q^t)(1-q)}(1-\varepsilon)^2 \right] \cdot R \right.$$
$$+ \left[ \frac{1-q^K}{1-q}(1-\varepsilon)^2 + \frac{q^K}{(1-q^t)(1-q)}\varepsilon^2 \right] \cdot P$$
$$\left. + \varepsilon(1-\varepsilon)(T+S) \right\}.$$

□

Regarding competition between various $AoN_K$ strategies, we can also employ comparable methods to obtain the corresponding payoffs and cooperation rates.

**Proposition 3.** *[$AoN_{Ka}vsAoN_{Kb}$] Consider the same game parameters as in Assumption 1. Suppose that two players now use different $AoN_K$ strategies, denoted as $AoN_{Ka}$ and $AoN_{Kb}$ respectively, and*

*that $Ka > kb$. Then the average cooperation rate is*

$$\rho(AoN_{Ka}, AoN_{Kb}) = \varepsilon x_0 \frac{1-q^{Kb}}{1-q} + \frac{1}{2}x_0 q^{Kb}[1-(1-q)^{Ka-Kb}] \tag{S5}$$
$$+(1-\varepsilon)x_0 q^{Kb}(1-q)^{Ka-Kb-1},$$

*where $q = \varepsilon^2 + (1-\varepsilon)^2$ and $x_0 = \left[\frac{1-q^{Kb}}{1-q} + q^{Kb-1} - q^{Kb-1}(1-q)^{Ka-Kb} + q^{Kb}(1+q)^{Ka-Kb-1}\right]^{-1}$. And the corresponding payoffs are*

$$\pi_{AoN_{Ka}} = \left[\frac{1-q^{Kb}}{1-q}\varepsilon^2 + (q^{Kb-1} - q^{Kb-1}(1-q)^{Ka-Kb})\varepsilon(1-\varepsilon) + q^{Kb}(1-q)^{Ka-Kb-1}(1-\varepsilon)^2\right]x_0 R$$
$$+\left[\frac{1-q^{Kb}}{1-q}\varepsilon(1-\varepsilon) + (q^{Kb-1} - q^{Kb-1}(1-q)^{Ka-Kb})(1-\varepsilon)^2 + q^{Kb}(1-q)^{Ka-Kb-1}\varepsilon(1-\varepsilon)\right]x_0 T$$
$$+\left[\frac{1-q^{Kb}}{1-q}\varepsilon(1-\varepsilon) + (q^{Kb-1} - q^{Kb-1}(1-q)^{Ka-Kb})\varepsilon^2 + q^{Kb}(1-q)^{Ka-Kb-1}\varepsilon(1-\varepsilon)\right]x_0 S$$
$$+\left[\frac{1-q^{Kb}}{1-q}(1-\varepsilon)^2 + (q^{Kb-1} - q^{Kb-1}(1-q)^{Ka-Kb})\varepsilon(1-\varepsilon) + q^{Kb}(1-q)^{Ka-Kb-1}\varepsilon^2\right]x_0 P \tag{S6}$$

*and*

$$\pi_{AoN_{Kb}} = \left[\frac{1-q^{Kb}}{1-q}\varepsilon^2 + (q^{Kb-1} - q^{Kb-1}(1-q)^{Ka-Kb})\varepsilon(1-\varepsilon) + q^{Kb}(1-q)^{Ka-Kb-1}(1-\varepsilon)^2\right]x_0 R$$
$$+\left[\frac{1-q^{Kb}}{1-q}\varepsilon(1-\varepsilon) + (q^{Kb-1} - q^{Kb-1}(1-q)^{Ka-Kb})\varepsilon^2 + q^{Kb}(1-q)^{Ka-Kb-1}\varepsilon(1-\varepsilon)\right]x_0 T$$
$$+\left[\frac{1-q^{Kb}}{1-q}\varepsilon(1-\varepsilon) + (q^{Kb-1} - q^{Kb-1}(1-q)^{Ka-Kb})(1-\varepsilon)^2 + q^{Kb}(1-q)^{Ka-Kb-1}\varepsilon(1-\varepsilon)\right]x_0 S$$
$$+\left[\frac{1-q^{Kb}}{1-q}(1-\varepsilon)^2 + (q^{Kb-1} - q^{Kb-1}(1-q)^{Ka-Kb})\varepsilon(1-\varepsilon) + q^{Kb}(1-q)^{Ka-Kb-1}\varepsilon^2\right]x_0 P, \tag{S7}$$

*respectively.*

*Proof.* Since $Ka > Kb$, the minimum number of states that completely describe the process of the game between the strategies is $Ka + 1$. We first need to calculate the value of $x_i$. According to the definition of $AoN_K$, we get

$$x_i = \begin{cases} qx_{i-1} & i = 1, \cdots, K_b \\ (1-q)x_{i-1} & i = K_b + 1, \cdots, K_a - 1 \\ (1-q)x_{i-1} + qx_i & i = K_a. \end{cases}$$

Because the sum of all $x_i$ is one, we get $x_0 = \left[\frac{1-q^{kb}}{1-q} + q^{kb-1} - q^{kb-1}(1-q)^{ka-kb} + q^{kb}(1+q)^{ka-kb-1}\right]^{-1}$.

5

|        | $AoN_1$ | $AoN_2$ | $AoN_3$ | $AoN_4$ | $AoN_5$ | $AoN_6$ | $AoN_7$ | $AoN_8$ | $AoN_9$ | $AoN_{10}$ |
|---|---|---|---|---|---|---|---|---|---|---|
| $AoN_1$    | 2.951 | 1.351 | 0.559 | 0.536 | 0.535 | 0.535 | 0.535 | 0.535 | 0.535 | 0.535 |
| $AoN_2$    | 2.974 | 2.913 | 1.269 | 0.717 | 0.703 | 0.702 | 0.702 | 0.702 | 0.702 | 0.702 |
| $AoN_3$    | 2.985 | 2.475 | 2.876 | 1.219 | 0.797 | 0.786 | 0.786 | 0.786 | 0.786 | 0.786 |
| $AoN_4$    | 2.985 | 2.329 | 2.176 | 2.839 | 1.186 | 0.845 | 0.837 | 0.837 | 0.837 | 0.837 |
| $AoN_5$    | 2.985 | 2.325 | 1.997 | 1.976 | 2.803 | 1.163 | 0.877 | 0.871 | 0.870 | 0.870 |
| $AoN_6$    | 2.985 | 2.325 | 1.993 | 1.798 | 1.833 | 2.768 | 1.145 | 0.900 | 0.895 | 0.895 |
| $AoN_7$    | 2.985 | 2.325 | 1.993 | 1.793 | 1.664 | 1.726 | 2.734 | 1.131 | 0.917 | 0.913 |
| $AoN_8$    | 2.985 | 2.325 | 1.993 | 1.793 | 1.660 | 1.568 | 1.642 | 2.700 | 1.120 | 0.931 |
| $AoN_9$    | 2.985 | 2.325 | 1.993 | 1.793 | 1.660 | 1.565 | 1.496 | 1.575 | 2.667 | 1.111 |
| $AoN_{10}$ | 2.985 | 2.325 | 1.993 | 1.793 | 1.660 | 1.565 | 1.493 | 1.441 | 1.521 | 2.635 |

Table 1: The payoff matrix of the game between different $AoN_K$ strategies.

Then, the cooperation rate of the two players becomes

$$P(AoN_{Ka}, AoN_{Kb}) = \sum_{i=0}^{Kb-1} x_i[\varepsilon^2 + \varepsilon(1-\varepsilon)] + \sum_{i=Kb}^{Ka-1} x_i[\varepsilon(1-\varepsilon) + \frac{1}{2}(\varepsilon^2 + (1-\varepsilon)^2)]$$
$$+ x_{Ka}[(1-\varepsilon)^2 + \varepsilon(1-\varepsilon)].$$

Finally, the payoffs of player $AoN_{Ka}$ and player $AoN_{Kb}$ become

$$\pi_{AoN_{Ka}} = \sum_{i=0}^{Kb-1} x_i[\varepsilon^2 R + (1-\varepsilon)^2 P + \varepsilon(1-\varepsilon)(T+S)]$$
$$+ \sum_{i=Kb}^{Ka-1} x_i[(1-\varepsilon)^2 T + \varepsilon^2 S + \varepsilon(1-\varepsilon)(P+R)]$$
$$+ x_{Ka}[(1-\varepsilon)^2 R + \varepsilon^2 P + \varepsilon(1-\varepsilon)(T+S)].$$

and

$$\pi_{AoN_{Kb}} = \sum_{i=0}^{Kb-1} x_i[\varepsilon^2 R + (1-\varepsilon)^2 P + \varepsilon(1-\varepsilon)(T+S)]$$
$$+ \sum_{i=Kb}^{Ka-1} x_i[(1-\varepsilon)^2 S + \varepsilon^2 T + \varepsilon(1-\varepsilon)(P+R)]$$
$$+ x_{Ka}[(1-\varepsilon)^2 R + \varepsilon^2 P + \varepsilon(1-\varepsilon)(T+S)].$$

Plugging $x_i$ into the equations, we will get the result shown by the proposition.

□

## 2 Competition with Unconditional Strategies

Next, we explore competition against two unconditional strategies ($ALLC$ and $ALLD$).

**Proposition 4.** *[AoN$_K$]* Consider an infinitely repeated prisoner's dilemma with error rate $\varepsilon > 0$ and make $q = \varepsilon^2 + (1-\varepsilon)^2$.

- *if one player uses AoN$_K$ strategy and the other uses ALLC, their payoffs are given by*

$$\pi(AoN_K, ALLC) = \left[\frac{1-(1-q)^K}{1-(1-2q)(1-q)^{K-1}}(1-\varepsilon)^2 + \frac{q(1-q)^{K-1}}{1-(1-2q)(1-q)^{K-1}}\varepsilon(1-\varepsilon)\right]T$$
$$+ \left[\frac{1-(1-q)^K}{1-(1-2q)(1-q)^{K-1}}\varepsilon(1-\varepsilon) + \frac{q(1-q)^{K-1}}{1-(1-2q)(1-q)^{K-1}}(1-\varepsilon)^2\right]R$$
$$+ \left[\frac{1-(1-q)^K}{1-(1-2q)(1-q)^{K-1}}\varepsilon(1-\varepsilon) + \frac{q(1-q)^{K-1}}{1-(1-2q)(1-q)^{K-1}}\varepsilon^2\right]P$$
$$+ \left[\frac{1-(1-q)^K}{1-(1-2q)(1-q)^{K-1}}\varepsilon^2 + \frac{q(1-q)^{K-1}}{1-(1-2q)(1-q)^{K-1}}\varepsilon(1-\varepsilon)\right]S$$
(S8)

*and*

$$\pi(ALLC, AoN_K) = \left[\frac{1-(1-q)^K}{1-(1-2q)(1-q)^{K-1}}\varepsilon^2 + \frac{q(1-q)^{K-1}}{1-(1-2q)(1-q)^{K-1}}\varepsilon(1-\varepsilon)\right]T$$
$$+ \left[\frac{1-(1-q)^K}{1-(1-2q)(1-q)^{K-1}}\varepsilon(1-\varepsilon) + \frac{q(1-q)^{K-1}}{1-(1-2q)(1-q)^{K-1}}(1-\varepsilon)^2\right]R$$
$$+ \left[\frac{1-(1-q)^K}{1-(1-2q)(1-q)^{K-1}}\varepsilon(1-\varepsilon) + \frac{q(1-q)^{K-1}}{1-(1-2q)(1-q)^{K-1}}\varepsilon^2\right]P$$
$$+ \left[\frac{1-(1-q)^K}{1-(1-2q)(1-q)^{K-1}}(1-\varepsilon)^2 + \frac{q(1-q)^{K-1}}{1-(1-2q)(1-q)^{K-1}}\varepsilon(1-\varepsilon)\right]S.$$
(S9)

- *if one player uses AoN$_K$ strategy and the other uses ALLD, their payoffs are given by*

$$\pi(AoN_K, ALLD) = \left[\frac{1-q^K}{1+q^{K-1}-2q^K}\varepsilon(1-\varepsilon) + \frac{(1-q)q^{K-1}}{1+q^{K-1}-2q^K}\varepsilon^2\right]T$$
$$+ \left[\frac{1-q^K}{1+q^{K-1}-2q^K}\varepsilon^2 + \frac{(1-q)q^{K-1}}{1+q^{K-1}-2q^K}\varepsilon(1-\varepsilon)\right]R$$
$$+ \left[\frac{1-q^K}{1+q^{K-1}-2q^K}(1-\varepsilon)^2 + \frac{(1-q)q^{K-1}}{1+q^{K-1}-2q^K}\varepsilon(1-\varepsilon)\right]P$$
$$+ \left[\frac{1-q^K}{1+q^{K-1}-2q^K}\varepsilon(1-\varepsilon) + \frac{(1-q)q^{K-1}}{1+q^{K-1}-2q^K}(1-\varepsilon)^2\right]S$$
(S10)

and

$$
\begin{aligned}
\pi(ALLD, AoN_K) = &\left[\frac{1-q^K}{1+q^{K-1}-2q^K}\varepsilon(1-\varepsilon) + \frac{(1-q)q^{K-1}}{1+q^{K-1}-2q^K}(1-\varepsilon)^2\right]T \\
&+ \left[\frac{1-q^K}{1+q^{K-1}-2q^K}\varepsilon^2 + \frac{(1-q)q^{K-1}}{1+q^{K-1}-2q^K}\varepsilon(1-\varepsilon)\right]R \\
&+ \left[\frac{1-q^K}{1+q^{K-1}-2q^K}(1-\varepsilon)^2 + \frac{(1-q)q^{K-1}}{1+q^{K-1}-2q^K}\varepsilon(1-\varepsilon)\right]P \\
&+ \left[\frac{1-q^K}{1+q^{K-1}-2q^K}\varepsilon(1-\varepsilon) + \frac{(1-q)q^{K-1}}{1+q^{K-1}-2q^K}\varepsilon^2\right]S.
\end{aligned} \quad (S11)
$$

*Proof.* The proof is similar to that of Proposition 1. □

As for $ADCO$ strategies with tolerance $t$, we have the following proposition.

**Proposition 5.** *Consider an infinitely repeated prisoner's dilemma with error rate $\varepsilon > 0$ and make $q = \varepsilon^2 + (1-\varepsilon)^2$.*

- *if one player uses the ADCO strategy with cooperation threshold $K$ and tolerance $t$ and the other uses ALLC, their payoffs are given by*

$$
\begin{aligned}
\pi(ADCO, ALLC) = &\ x_0 \left[\frac{1-(1-q)^K}{q}(1-\varepsilon)^2 + \frac{(1-q)^{K-1}}{1-q^t}\varepsilon(1-\varepsilon)\right]T \\
&+ x_0 \left[\frac{1-(1-q)^K}{q}\varepsilon(1-\varepsilon) + \frac{(1-q)^{K-1}}{1-q^t}(1-\varepsilon)^2\right]R \\
&+ x_0 \left[\frac{1-(1-q)^K}{q}\varepsilon(1-\varepsilon) + \frac{(1-q)^{K-1}}{1-q^t}\varepsilon^2\right]P \\
&+ x_0 \left[\frac{1-(1-q)^K}{q}\varepsilon^2 + \frac{(1-q)^{K-1}}{1-q^t}\varepsilon(1-\varepsilon)\right]S
\end{aligned} \quad (S12)
$$

*and*

$$
\begin{aligned}
\pi(ALLC, ADCO) = &\ x_0 \left[\frac{1-(1-q)^K}{q}\varepsilon^2 + \frac{(1-q)^{K-1}}{1-q^t}\varepsilon(1-\varepsilon)\right]T \\
&+ x_0 \left[\frac{1-(1-q)^K}{q}\varepsilon(1-\varepsilon) + \frac{(1-q)^{K-1}}{1-q^t}(1-\varepsilon)^2\right]R \\
&+ x_0 \left[\frac{1-(1-q)^K}{q}\varepsilon(1-\varepsilon) + \frac{(1-q)^{K-1}}{1-q^t}\varepsilon^2\right]P \\
&+ x_0 \left[\frac{1-(1-q)^K}{q}(1-\varepsilon)^2 + \frac{(1-q)^{K-1}}{1-q^t}\varepsilon(1-\varepsilon)\right]S,
\end{aligned} \quad (S13)
$$

where $x_0 = \left[\frac{1-(1-q)^K}{q} + \frac{(1-q)^{K-1}}{1-q^t}\right]^{-1}$.

- *if one player uses ADCO strategy with parameters $K$ and $t$ and the other uses ALLD, their*

8

*payoffs are given by*

$$\pi(ADCO, ALLD) = x_0 \left[\frac{1-q^K}{1-q}\varepsilon(1-\varepsilon) + \frac{q^{K-1}}{1-(1-q)^t}\varepsilon^2\right] T$$
$$+ x_0 \left[\frac{1-q^K}{1-q}\varepsilon^2 + \frac{q^{K-1}}{1-(1-q)^t}\varepsilon(1-\varepsilon)\right] R$$
$$+ x_0 \left[\frac{1-q^K}{1-q}(1-\varepsilon)^2 + \frac{q^{K-1}}{1-(1-q)^t}\varepsilon(1-\varepsilon)\right] P$$
$$+ x_0 \left[\frac{1-q^K}{1-q}\varepsilon(1-\varepsilon) + \frac{q^{K-1}}{1-(1-q)^t}(1-\varepsilon)^2\right] S$$

(S14)

*and*

$$\pi(ALLD, ADCO) = x_0 \left[\frac{1-q^K}{1-q}\varepsilon(1-\varepsilon) + \frac{q^{K-1}}{1-(1-q)^t}(1-\varepsilon)^2\right] T$$
$$+ x_0 \left[\frac{1-q^K}{1-q}\varepsilon^2 + \frac{q^{K-1}}{1-(1-q)^t}\varepsilon(1-\varepsilon)\right] R$$
$$+ x_0 \left[\frac{1-q^K}{1-q}(1-\varepsilon)^2 + \frac{q^{K-1}}{1-(1-q)^t}\varepsilon(1-\varepsilon)\right] P$$
$$+ x_0 \left[\frac{1-q^K}{1-q}\varepsilon(1-\varepsilon) + \frac{q^{K-1}}{1-(1-q)^t}\varepsilon^2\right] S,$$

(S15)

*where* $x_0 = \left[\frac{1-q^K}{1-q} + \frac{q^{K-1}}{1-(1-q)^t}\right]^{-1}$.

*Proof.* The proof is similar to that of Proposition 1. □

## 3 Competition with arbitrary Memory-1 Strategies

Next, let's consider the case where Player 1 adopts an $AoN_K$ (or $ADCO$) strategy, while Player 2 employs an arbitrary memory-1 strategy denoted as $p = [p_{CC}, p_{CD}, p_{DC}, p_{DD}]$. In this scenario, we examine the infinitely repeated Prisoner's Dilemma with an error rate of $\varepsilon$. Incorporating implementation error into the strategy, the memory-1 strategy becomes $p' = [p'_{CC}, p'_{CD}, p'_{DC}, p'_{DD}] = (1-\varepsilon)p + \varepsilon(1-p)$. The decision of the memory-1 strategy relies on the outcomes of the most recent round. So, we then add the information of the most recent round to the $A_i$ state, denoted as $\{A_0\}^b_a$ where Player 1's action being $a$ and Player 2's action being $b$ in the latest round. In state $A_0$, the memory-1 strategy needs to specify whether it is $\{A_0\}^D_C$ or $\{A_0\}^C_D$. Let $\{x_i\}^b_a$ represent the probability of the player being in a state $\{A_i\}^b_a$ ($i = 0, 1, 2, \cdots, K$) or $\{A_{K,i}\}^b_a$ ($i = K+1, K+2, \cdots, K+t$). Here, $a$ and $b$ can take the actions of either "C" (cooperate) or "D" (defect). Thus, for the repeated interactions between $AoN_K$ and memory-1 strategies, we have $2K+2$ state probabilities: $\{x_0\}^D_C$, $\{x_0\}^C_D$, $\{x_1\}^C_C$, $\{x_1\}^D_D$, and so on up to $\{x_K\}^C_C$ and $\{x_K\}^D_D$. Since errors are present, all states become ergodic. Similar to

the previous analysis, We can establish a system of equations for $\{x_i\}_a^b$, given by

$$x_{0C}^D = (1-\varepsilon)p'_{DC} \cdot x_{0D}^C + (1-\varepsilon)p'_{CD} \cdot x_{0C}^D + (1-\varepsilon)p'_{CC}(x_{1C}^C + x_{2C}^C + \cdots + x_{K-1C}^C)$$
$$+ (1-\varepsilon)p'_{DD}(x_{1D}^D + x_{2D}^D + \cdots + x_{K-1D}^D) + \varepsilon p'_{CC}(x_{KC}^C + x_{K+1C}^C + \cdots + x_{IC}^C)$$
$$+ \varepsilon p'_{DD}(x_{KD}^D + x_{K+1D}^D + \cdots + x_{ID}^D)$$

$$x_{1C}^C = \varepsilon p'_{DC} \cdot x_{0D}^C + \varepsilon p'_{CD} \cdot x_{0C}^D \qquad x_{1D}^D = (1-\varepsilon)(1-p'_{DC})x_{0D}^C + (1-\varepsilon)(1-p'_{CD})x_{0C}^D$$
$$x_{2C}^C = \varepsilon p'_{CC} \cdot x_{1C}^C + \varepsilon p'_{DD} \cdot x_{1D}^D \qquad x_{2D}^D = (1-\varepsilon)(1-p'_{CC})x_{1C}^C + (1-\varepsilon)(1-p'_{DD})x_{1D}^D$$
$$\vdots \qquad\qquad\qquad\qquad \vdots$$
$$x_{K-1C}^C = (1-\varepsilon)p'_{CC} \cdot x_{K-2C}^C + (1-\varepsilon)p'_{DD} \cdot x_{K-2D}^D \quad x_{K-1D}^D = \varepsilon(1-p'_{CC})x_{K-2C}^C + \varepsilon(1-p'_{DD})x_{K-2D}^D$$

$$x_{KC}^C = (1-\varepsilon)p'_{CC} \cdot x_{K-1C}^C + (1-\varepsilon)p'_{DD} \cdot x_{K-1D}^D + (1-\varepsilon)p'_{CC} \cdot x_{KC}^C + (1-\varepsilon)p'_{DD} \cdot x_{KD}^D$$
$$x_{KD}^D = \varepsilon(1-p'_{CC})x_{K-1C}^C + \varepsilon(1-p'_{DD})x_{K-1D}^D + \varepsilon(1-p'_{CC})x_{KC}^C + \varepsilon(1-p'_{DD})x_{KD}^D$$

The above yields $2K+1$ equations. Combined with the equation $\sum_{0\leq i\leq K}\sum_{a,b} x_{ia}^b = 1$, we can solve the system of equations to determine the value of $\{x_i\}_a^b$. After that, we can derive the payoffs of the two players as

$$\pi_{AoN_K} = x_{0C}^D T + x_{0D}^C S + \sum_{1\leq i\leq I} x_{iC}^C R + \sum_{1\leq i\leq I} x_{iD}^D P$$
$$\pi_{Memory-1} = x_{0C}^D S + x_{0D}^C T + \sum_{1\leq i\leq I} x_{iC}^C R + \sum_{1\leq i\leq I} x_{iD}^D P,$$

respectively.

The payoff calculation method between $ADCO$ with cooperation threshold $K$ and tolerance $t$ and the memory-1 strategy is similar to that for $AoN_K$. We incorporate two state probabilities, $\{x_K\}_C^D$ and $\{x_K\}_D^C$, in accordance with the tolerance properties of the strategy. There are $2(K+t)+4$ states. The system of equations is given as follows:

$$x_{0C}^D = (1-\varepsilon)p'_{DC} \cdot x_{0D}^C + (1-\varepsilon)p'_{CD} \cdot x_{0C}^D + (1-\varepsilon)p'_{CC}(x_{1C}^C + x_{2C}^C + \cdots + x_{K-1C}^C)$$
$$+ (1-\varepsilon)p'_{DD}(x_{1D}^D + x_{2D}^D + \cdots + x_{K-1D}^D) + \varepsilon p'_{CC}(x_{KC}^C + x_{K+1C}^C + \cdots + x_{K+t-1C}^C)$$
$$+ \varepsilon p'_{DD}(x_{KD}^D + x_{K+1D}^D + \cdots + x_{K+t-1D}^D) + \varepsilon p'_{DC} \cdot x_{KD}^C + \varepsilon p'_{CD} \cdot x_{KC}^D$$

$$x_{1C}^C = \varepsilon p'_{DC} \cdot x_{0D}^C + \varepsilon p'_{CD} \cdot x_{0C}^D \qquad x_{1D}^D = (1-\varepsilon)(1-p'_{DC})x_{0D}^C + (1-\varepsilon)(1-p'_{CD})x_{0C}^D$$
$$x_{2C}^C = \varepsilon p'_{CC} \cdot x_{1C}^C + \varepsilon p'_{DD} \cdot x_{1D}^D \qquad x_{2D}^D = (1-\varepsilon)(1-p'_{CC})x_{1C}^C + (1-\varepsilon)(1-p'_{DD})x_{1D}^D$$
$$\vdots \qquad\qquad\qquad\qquad \vdots$$
$$x_{K+1C}^C = (1-\varepsilon)p'_{CC} \cdot x_{KC}^C + (1-\varepsilon)p'_{DD} \cdot x_{KD}^D + (1-\varepsilon)p'_{DC} \cdot x_{KD}^C + (1-\varepsilon)p'_{CD} \cdot x_{KC}^D$$
$$x_{K+1D}^D = \varepsilon(1-p'_{CC}) \cdot x_{KC}^C + \varepsilon(1-p'_{DD}) \cdot x_{KD}^D + \varepsilon(1-p'_{DC}) \cdot x_{KD}^C + \varepsilon(1-p'_{CD}) \cdot x_{KC}^D$$

10

$$x_{K+2\,C}^{C} = (1-\varepsilon)p'_{CC} \cdot x_{K+1\,C}^{C} + (1-\varepsilon)p'_{DD} \cdot x_{K+1\,D}^{D} \quad x_{K+2\,D}^{D} = \varepsilon(1-p'_{CC})x_{K+1\,C}^{C} + \varepsilon(1-p'_{DD})x_{K+1\,D}^{D}$$
$$\vdots \qquad\qquad \vdots$$
$$x_{K+t\,C}^{C} = (1-\varepsilon)p'_{CC} \cdot x_{K+t-1\,C}^{C} + (1-\varepsilon)p'_{DD} \cdot x_{K+t-1\,D}^{D} + (1-\varepsilon)p'_{CC} \cdot x_{K+t\,C}^{C} + (1-\varepsilon)p'_{DD} \cdot x_{K+t\,D}^{D}$$
$$x_{K+t\,D}^{D} = \varepsilon(1-p'_{CC}) \cdot x_{K+t-1\,C}^{C} + \varepsilon(1-p'_{DD}) \cdot x_{K+t-1\,D}^{D} + \varepsilon(1-p'_{CC}) \cdot x_{K+t\,C}^{C} + \varepsilon(1-p'_{DD}) \cdot x_{K+t\,D}^{D}.$$
$$x_{K\,D}^{C} = (1-\varepsilon)(1-p'_{CC}) \cdot x_{K+t\,C}^{C} + (1-\varepsilon)(1-p'_{DD}) \cdot x_{K+t\,D}^{D} \quad x_{K\,C}^{D} = \varepsilon p'_{CC} \cdot x_{K+t\,C}^{C} + \varepsilon p'_{DD} \cdot x_{K+t\,D}^{D}.$$

Combined with the equation $\sum_{0 \leq i \leq K} \sum_{a,b} x_{i\,a}^{b} = 1$, there are $2(K+t)+4$ equations. We can determine the value of $x_{i\,a}^{b}$ by solving a system of equations. Thus, the players' corresponding payoffs are

$$\pi_{ADCO} = (x_{0\,C}^{D} + x_{K\,C}^{D})T + (x_{0\,D}^{C} + x_{K\,D}^{C})S + \sum_{1 \leq i \leq K+t} x_{i\,C}^{C} R + \sum_{1 \leq i \leq K+t} x_{i\,D}^{D} P$$

$$\pi_{Memory-1} = (x_{0\,C}^{D} + x_{K\,C}^{D})S + (x_{0\,D}^{C} + x_{K\,D}^{C})T + \sum_{1 \leq i \leq K+t} x_{i\,C}^{C} R + \sum_{1 \leq i \leq K+t} x_{i\,D}^{D} P.$$

We computed the payoff between $ADCO$ and any memory-1 strategy by solving a system of linear equations. In the main text, we selected ten representative classic strategies known for their performance in various competitive environments. These include two unconditional strategies and three pure strategies: $WSLS$ ([1,0,0,1]), $Grim$([1,0,0,0]), $TFT$([1,0,1,0]), as well as five stochastic strategies: $GTFT_{0.2}$([1,0.2,1,0.2]), $GTFT_{0.4}$([1,0.4,1,0.4]), $ZD_2$([8/9,1/2,1/3,0]), $ZD_3$([11/13,1/2,7/26,0]), $Random$([0.5,0.5,0.5,0.5]).

## 4 Competition with some strategies through computer simulation

Furthermore, in addition to traditional memory-based strategies, we also investigate strategies that go beyond the constraints of finite memory. Without sacrificing generality, we examine two distinct types of strategies: 'HardMajority' and 'Cumulative Reciprocity'. In the 'HardMajority' strategy, a player defects in the initial round and continues to do so unless the opponent cooperates more times than defects. On the other hand, in the 'Cumulative Reciprocity' strategy, a player defects only if the opponent's count of defections surpasses the player's count of defections plus $\Delta$; otherwise, the player cooperates. To analyze the outcomes, we employ computer simulations to derive individual payoffs.

## 5 Evolutionary game dynamics

Nash equilibrium is a static property of games between strategies. Instead, evolutionary game theory provides a classic paradigm for studying population games in which individuals can imitate or learn. More specifically, in an evolutionary system involving many players, players adopt a strategy to interact with other individuals and obtain benefits from the interactions. Successful strategies (with higher payoffs) are more likely to be imitated and diffuse in the population. This paper adopts two

evolutionary methods for two-strategy and multi-strategy competition respectively. For the pairwise competition between $ADCO$ (or $AoN_K$) strategy and one other strategy in the population, we consider the large population and pay attention to the evolution process of the strategies. For the evolution of multiple strategies in the population, we focus on the abundance of each strategy in the evolutionary competition.

## Description of population evolution with two strategies

Consider a large, well-mixed population with two strategies, denoted as strategy $A$ and strategy $B$. $x_A$ represents the frequency of strategy $A$ in the population. So $x_B = 1 - x_A$. The fitness of strategy $A$ is defined as the average payoff of a strategy $A$ player playing against the entire population of players, that is $f(A) = x_A \pi(A, B) + x_B \pi(B, A)$. Similarly, $f(B) = x_A \pi(B, A) + x_B \pi(A, B)$. The average fitness of the population is written as $\bar{f} = x_A f(A) + x_B f(B)$. Next, players will evolve based on their fitness. The frequency of strategy $A$ in the next generation is determined to be $x'_A = x_A f(A) / \bar{f}$. This basic evolutionary step is repeated until the frequency of each strategy does not change. Either the population is stuck with only one strategy, or both strategies coexist.

## Description of multi-strategy population evolution

Different from the above, we here consider a finite population of size M. Each pair of players interacts and participates in a repeated game. In each generation, a player is randomly selected from the population. With probability $\mu$ (the mutation has occurred), this player will randomly select a strategy from the strategy set to become his new strategy. With probability $1 - \mu$, the player will compare his payoff $\pi$ with the payoff $\tilde{\pi}$ of a randomly selected member of the population. Assume that the focal player will switch to the other player's strategy with a probability

$$p = \frac{1}{1 + exp[-\beta(\tilde{\pi} - \pi)]},$$

where $\beta$ is the strength of selection. It describes the effect of payoff on strategy learning. $p = 0.5$ holds regardless of the difference in payoffs if $\beta = 0$. If $\beta \to \infty$, imitation always occurs despite the small difference in payoffs. This process will be repeated a sufficient number of times.

Here, we follow the framework of Imhof and Nowak and assume mutation is rare. It implies that the mutant either reaches the fixation in the population or goes extinct before the next mutant emerges. Therefore, the population is almost always homogeneous, where members all adopt the same strategy. The fixation probability of the mutant can be calculated to be

$$\rho_M = \left(1 + \sum_{i=1}^{N-1} \prod_{j=1}^{i} \exp -\beta \left[\pi_M(j) - \pi_R(j)\right]\right)^{-1}.$$

where $\pi_M(j)$ and $\pi_R(j)$ are the payoffs of the mutation strategy and the resident strategy respectively when there are $j$ mutants in the population. $N$ is the population size. Once the mutant reaches fixation or goes extinct, we introduce a new mutant into the resident population. We will repeat this elementary step sufficiently many times. The population will experience different strategies of occupation in the process, as different mutants invade again and again. However, different strategies occupy different proportions, and the evolution of robust strategies will take more time in this process. It is a Markov process that the population switches from one strategy to another. By calculating the corresponding stationary distribution, we can get the abundance of the strategies in the evolution process, just as described in the main text. However, if the mutation rate is large, this method is not feasible. Instead, we resort to computer simulations, iterating the elementary population updating process with $10^9$ mutant strategies per simulation run. This approach offers a pragmatic approximation for the dynamics.

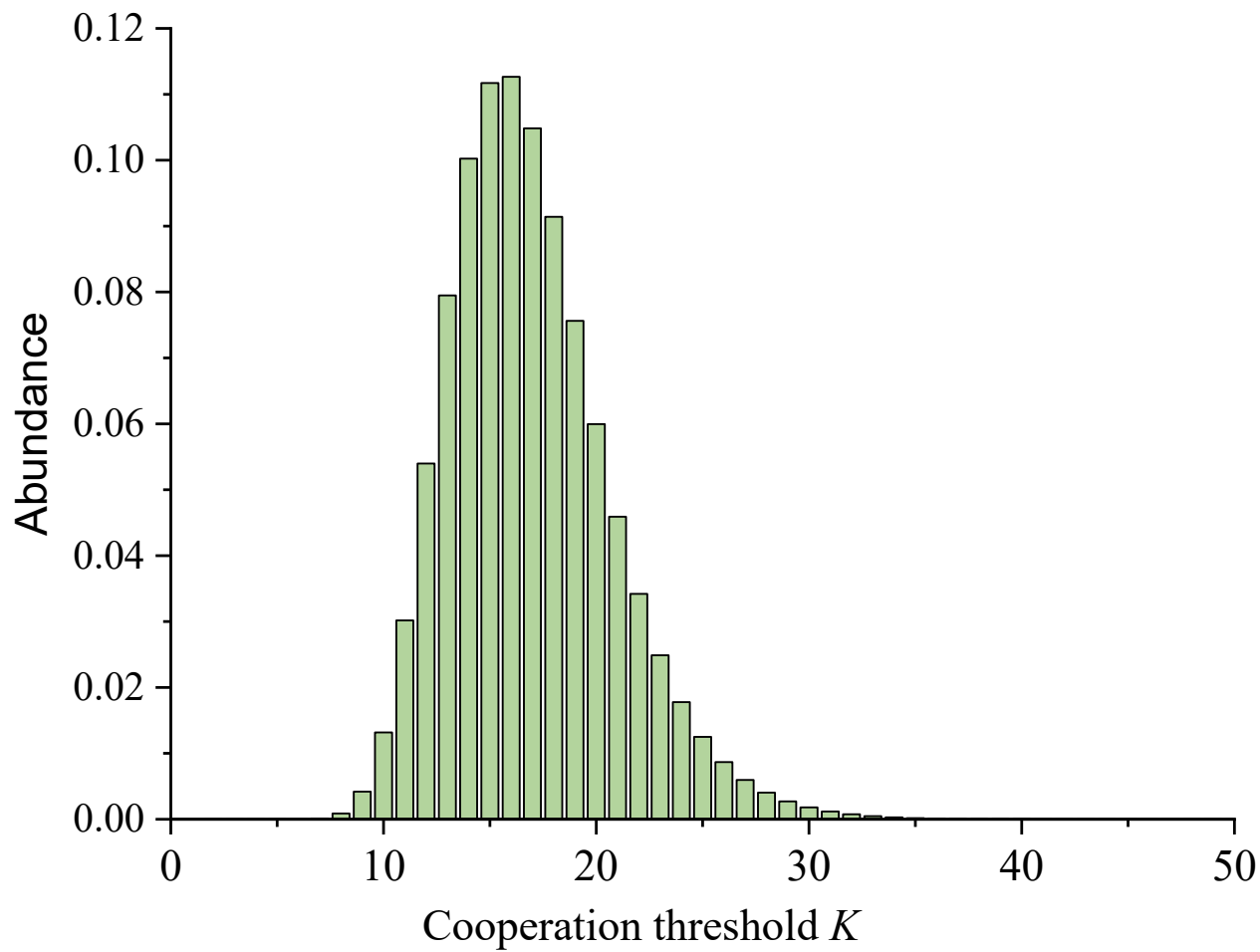

**Fig. S1.** Evolutionary competition between $AoN_K$ strategy under high mutation rates. We consider a sufficiently large class of 50 strategies ($K = 1, 2, \cdots, 50$) with mutation rates set $\mu = 0.1$. The simulation results show that the $AoN_{17}$ strategy in the middle has the highest abundance, and the strategy abundance on both sides gradually decreases. Similar to the results for rare mutations, the strategy abundance always peaks at a moderate memory length. Other parameters include a population size of $M = 100$, $\varepsilon = 0.01$, and selection strength of $\beta = 1$; The simulation is executed for $10^9$ mutant strategies.

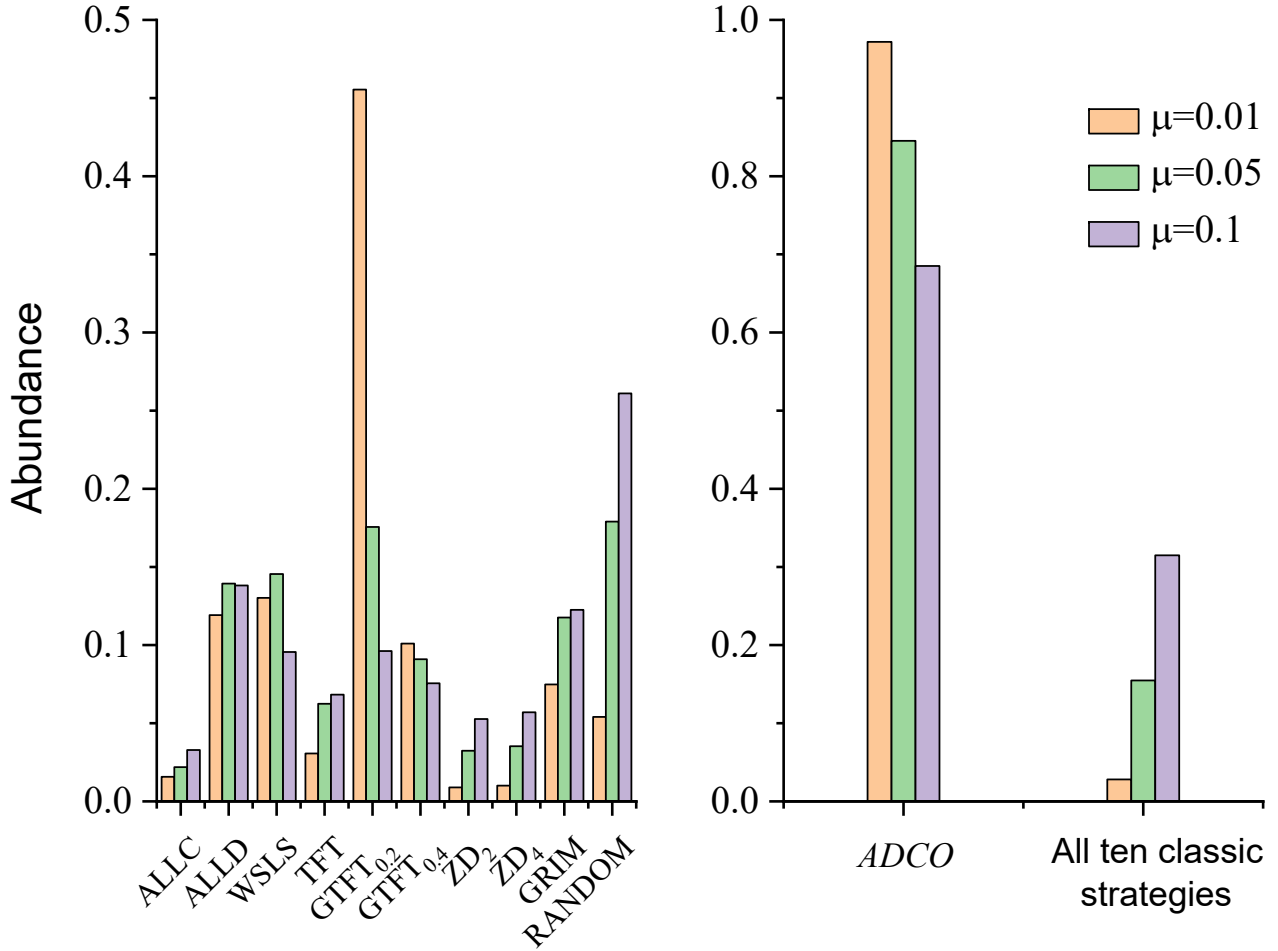

**Fig. S2.** Evolutionary competition of $ADCO$ with ten classical strategies under diverse mutation rates. (A) Three mutation rates are examined: $\mu = 0.01$, $\mu = 0.05$, $\mu = 0.1$. we first let these ten classic strategies compete in the evolutionary race. Through computer simulations, we observe that the most prevalent strategy in the population varies with mutation rates, with the $GTFT_{0.2}$ strategy dominating at relatively lower mutation rates ($\mu = 0.01$), while $RANDOM$ dominates at higher mutation rate $\mu = 0.05$ and $\mu = 0.1$. (B) Following the introduction of $ADCO(K = 3, t = 2)$, it almost entirely dominates the entire population, regardless of whether the mutation rate is high or low. Other parameters include a population size of $M = 100$, $\varepsilon = 0.01$, and selection strength of $\beta = 1$; each simulation is executed for $5 \times 10^8$ mutant strategies.

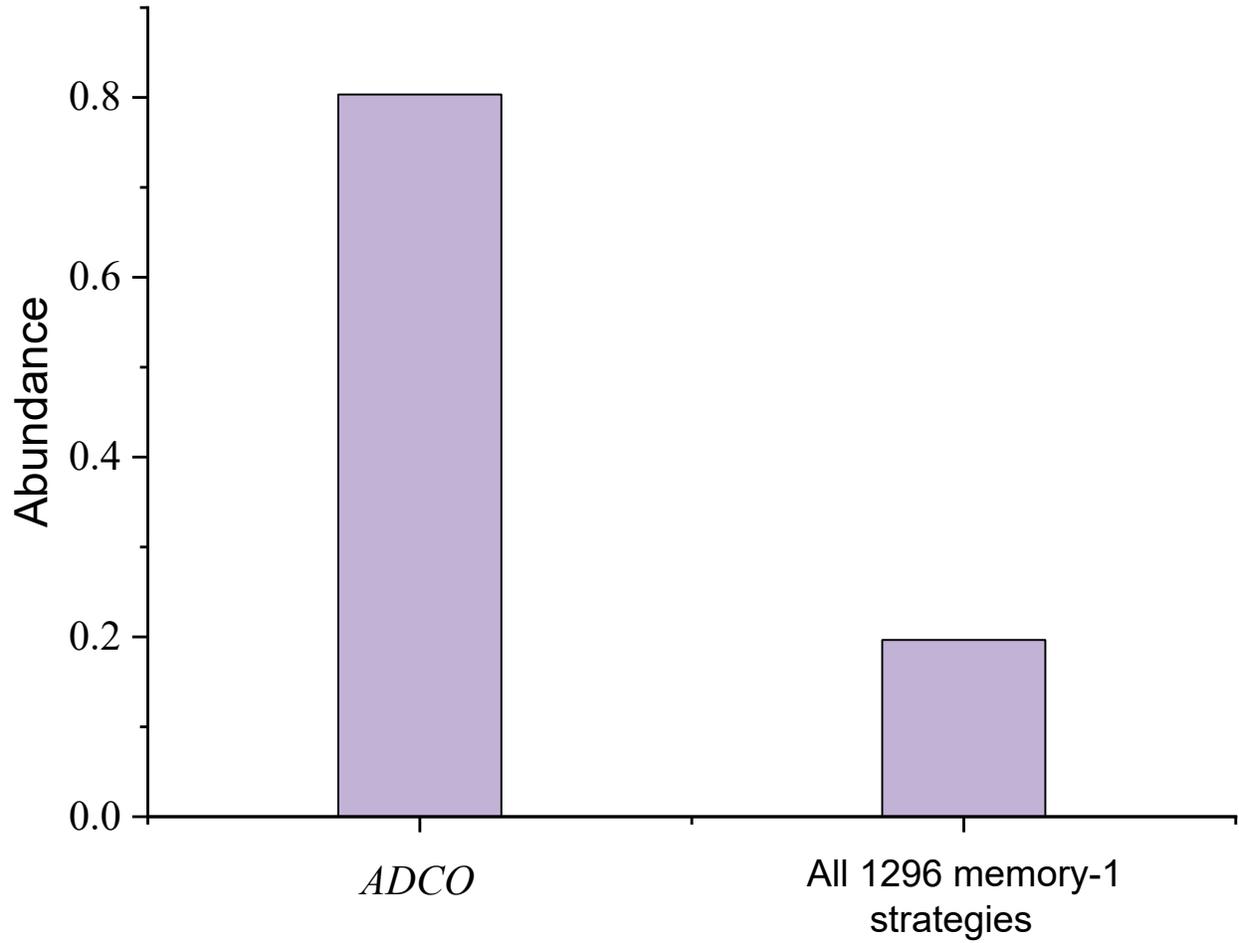

**Fig. S3.** Evolution of $ADCO$ in the population of memory-1 players under high mutation rates. We considered 1296 memory-1 strategies as described in the main text. Through simulations, we observed that, even at high mutation rates ($\mu = 0.1$), $ADCO(K = 3, t = 2)$ still becomes predominant, maintaining an average proportion close to 80% throughout the entire evolutionary process time steps in the population. Other parameters include a population size of $M = 100$, $\varepsilon = 0.01$, and selection strength of $\beta = 1$; each simulation is executed for $10^9$ mutant strategies.



\end{document}